# Prediction and observation of the first antiferromagnetic topological insulator

M. M. Otrokov,[1,2,3,4] I. I. Klimovskikh,[4] H. Bentmann,[5] D. Estyunin,[4] A. Zeugner,[6] Z. S. Aliev,[7,8] S. Gaß,[9] A. U. B. Wolter,[9] A. V. Koroleva,[4] A. M. Shikin,[4] M. Blanco-Rey,[10,3] M. Hoffmann,[11] I. P. Rusinov,[12,4] A. Yu. Vyazovskaya,[12,4] S. V. Eremeev,[13,12] Yu. M. Koroteev,[13,12] V. M. Kuznetsov,[12] F. Freyse,[14] J. Sánchez-Barriga,[14] I. R. Amiraslanov,[7] M. B. Babanly,[15] N. T. Mamedov,[7] N. A. Abdullayev,[7] V. N. Zverev,[16] A. Alfonsov,[9] V. Kataev,[9] B. Büchner,[9,17] E. F. Schwier,[18] S. Kumar,[18] A. Kimura,[19] L. Petaccia,[20] G. Di Santo,[20] R. C. Vidal,[5] S. Schatz,[5] K. Kißner,[5] M. Ünzelmann,[5] C. H. Min,[5] Simon K. Moser,[21] T. R. F. Peixoto,[5] F. Reinert,[5] A. Ernst,[11,22] P. M. Echenique,[3,10,1] A. Isaeva,[9,17] and E. V. Chulkov[10,3,1,4]

[1]Centro de Física de Materiales (CFM-MPC), Centro Mixto CSIC-UPV/EHU, 20018 Donostia-San Sebastián, Basque Country, Spain

[2]IKERBASQUE, Basque Foundation for Science, 48011 Bilbao, Spain

[3]Donostia International Physics Center (DIPC), 20018 Donostia-San Sebastián, Basque Country, Spain

[4]Saint Petersburg State University, 198504 Saint Petersburg, Russia

[5]Experimentelle Physik VII, Universität Wüurzburg, Am Hubland, D-97074 Würzburg, Germany

[6]Technische Universität Dresden, Faculty of Chemistry and Food Chemistry, Bergstrasse 66, Dresden, D-01069, Germany

[7]Institute of Physics, Azerbaijan National Academy of Sciences, 1143 Baku, Azerbaijan

[8]Azerbaijan State Oil and Industry University, AZ1010 Baku, Azerbaijan

[9]Leibniz IFW Dresden, Institute for Solid State Research, Helmholtzstrasse 20, D-01069, Germany

[10]Departamento de Física de Materiales UPV/EHU, 20080 Donostia-San Sebastián, Basque Country, Spain

[11]Institut für Theoretische Physik, Johannes Kepler Universität, A 4040 Linz, Austria

[12]Tomsk State University, 634050 Tomsk, Russia

[13]Institute of Strength Physics and Materials Science, Russian Academy of Sciences, 634021 Tomsk, Russia

[14]Helmholtz-Zentrum Berlin für Materialien und Energie, Elektronenspeicherring BESSY II, Albert-Einstein-Straße 15, 12489 Berlin, Germany

[15]Institute of Catalysis and Inorganic Chemistry, Azerbaijan National Academy of Science, AZ1143 Baku, Azerbaijan

[16]Institute of Solid State Physics, Russian Academy of Sciences, Chernogolovka 142432, Russia

[17]Technische Universität Dresden, Faculty of Physics, Haeckelstrasse 3, D-01069, Germany

[18]Hiroshima Synchrotron Radiation Center, Hiroshima University, Higashi-Hiroshima, Hiroshima 739-0046, Japan

[19]Department of Physical Sciences, Graduate School of Science, Hiroshima University, Higashi-Hiroshima, Hiroshima 739-8526, Japan

[20]Elettra Sincrotrone Trieste, Strada Statale 14 km 163.5, 34149 Trieste, Italy

[21]Advanced Light Source, Lawrence Berkeley National Laboratory, Berkeley, CA 94720, USA

[22]Max-Planck-Institut für Mikrostrukturphysik, Weinberg 2, D-06120 Halle, Germany

Corresponding authors: mikhail.otrokov@gmail.com, evguenivladimirovich.tchoulkov@ehu.eus

**Magnetic topological insulators (MTIs) are narrow gap semiconductor materials that combine non-trivial band topology and magnetic order[1]. Unlike their nonmagnetic counterparts, MTIs may have some of the surfaces gapped due to breaking the time-reversal symmetry, which enables a number of exotic phenomena having potential applications in spintronics[1-3]. So far, MTIs have only been created by means of doping nonmagnetic TIs with 3*d* transition metal elements, however, such an approach leads to strongly inhomogeneous magnetic[4] and electronic[5] properties of these materials, restricting the observation of important effects to very low temperatures[2,3]. Finding intrinsic MTI, i.e. a stoichiometric well-ordered magnetic compound, could be an ideal solution to these problems, but no such material was observed to date. Here, using density functional theory we predict and further confirm by means of structural, transport, magnetic, angle- and spin-resolved photoemission spectroscopy measurements the realization of the antiferromagnetic (AFM) TI phase, that is hosted by the van der Waals layered compound $MnBi_2Te_4$. An interlayer AFM ordering makes $MnBi_2Te_4$ invariant with respect to the combination of the time-reversal ($\Theta$) and primitive-lattice translation ($T_{1/2}$) symmetries, $S = \Theta T_{1/2}$, giving rise to the $Z_2$ topological classification of AFM insulators[6]. We find $Z_2 = 1$ for $MnBi_2Te_4$, which confirms its topologically nontrivial nature. The *S*-breaking (0001) surface of $MnBi_2Te_4$ exhibits a giant bandgap in the topological surface state as evidenced by *ab initio* calculations and photoemission measurements. These results culminate almost a decade-long search of an AFMTI,**

**predicted in 2010[6]. Furthermore, $MnBi_2Te_4$ is the first intrinsic magnetic TI realized experimentally. As an outcome of this, a number of fundamental phenomena are expected to eventually be observed, being among them quantized magnetoelectric coupling[6-8] and axion electrodynamics[9,10]. Other exotic phenomena could become accessible at temperatures significantly higher than those achieved to date, like the quantum anomalous Hall effect[2] and chiral Majorana fermions[3].**

The first reference to $MnBi_2Te_4$ as a stable chemical compound dates back to 2013 when it was synthesized in the powder form[11]. The trigonal structure (space group $R\bar{3}m$) of $MnBi_2Te_4$ comprises of septuple layer (SL) blocks stacked along the [0001] direction and bound to each other by van der Waals forces (Fig. 1a). High-quality $MnBi_2Te_4$ single crystals were for the first time grown at the TU Dresden (hereinafter D samples, Fig. 2a) and at IP Baku (B samples, Fig. 2b). Our single-crystal X-ray diffraction experiments confirm exactly the same lattice symmetry as reported in Ref. [11]. Since the magnetism of $MnBi_2Te_4$ was not experimentally investigated, we begin our study by calculating the exchange coupling parameters $J$ from first principles (Figs. 1b and 1c). Among the intralayer interactions $J^{\parallel}$ the one between the first nearest neighbors in the Mn layer is clearly dominant ($J_{0,1}^{||} \approx 0.09$ meV/$\mu_B^2$), while the interactions with more distant neighbors are an order of magnitude weaker, whereupon an FM ordering is expected within each SL block of $MnBi_2Te_4$. In contrast, the interlayer coupling constants $J^{\perp}$ are mostly negative resulting in a negative $J_0^{\perp} \approx \sum_{j\neq 0} J_{0,j}^{\perp}$ coefficient equal to –0.022 meV/$\mu_B^2$, which means that the overall coupling between neighboring Mn layers is antiferromagnetic (Fig. 1d). The magnetic anisotropy energy is positive and equal to 0.225 meV per Mn atom, indicating the easy axis with an out-of-plane orientation of the local magnetic moments (LMMs) of $\pm 4.607\mu_B$. Monte Carlo simulations confirm the interlayer AFM structure with a Néel temperature ($T_N$) of 3D ordering of 25.4 K (Extended Data Fig. 1).

Given the magnetic ground state, the bulk electronic structure was calculated. As shown in Fig. 1e, the system is insulating, the fundamental bandgap value, determined from the GGA+$U$ calculation of the density of states (DOS), is of ~220 meV. To determine whether the gap is negative (inverted), we performed the DOS calculations decreasing the spin-orbit coupling (SOC) constant $\lambda$ stepwise from its natural value $\lambda_0$ to $\lambda = 0.5\lambda_0$. It was found that at $\lambda/\lambda_0 \approx 0.675$ the gap is closed, while at other $\lambda/\lambda_0$ ratios it is nonzero, which points towards a nontrivial topology of $MnBi_2Te_4$.

Mong, Essin, and Moore[6] introduced a $Z_2$ classification of AFM insulators, provided by the $S = \Theta T_{1/2}$-symmetry, which is inherent to $MnBi_2Te_4$. We find $Z_2 = 1$, which classifies $MnBi_2Te_4$ as an AFMTI.

The implication of the bulk bandgap inversion in a TI is seen at its surface, where the topological phase manifests itself by the appearance of the topological surface state. In the case of nonmagnetic TIs, this surface state is gapless[12], however at the (0001) surface of the AFMTI $MnBi_2Te_4$ we find a 88-meV-wide bandgap (Fig. 1f). Here, the $S$-symmetry is broken and the out-of-plane magnetization of the near-surface FM layer opens the Dirac point (DP) gap[6]. In contrast, the $S$-preserving surface is gapless as it should be for an AFMTI (inset to Fig. 1f and Extended Data Fig. 2).

Temperature- and field-dependent magnetization measurements performed on D samples (Fig. 2c,d) establish a 3D AFM order below $T_N$ = 24.2(5) K, in agreement with the prediction by Monte Carlo method and results of the resistivity measurements (Extended Data Fig. 3a). Below $T_N$, a strongly anisotropic magnetic susceptibility $\chi$ is observed, that decreases much steeper for $H \perp (0001)$. No splitting between zero-field-cooled and field-cooled-warming curves was found. The paramagnetic (PM) regime above $T_N$ was fitted with a modified Curie-Weiss law, $\chi(T) = \chi_0 + C/(T - \Theta_{CW})$, in the 100 K to 250 K range. Here, $\chi_0$ stands for a temperature-independent magnetic susceptibility of both diamagnetic closed electron shells and a Pauli PM contribution due to some degree of metallicity in this material (see below). $C/(T - \Theta_{CW})$ considers a temperature-dependent Curie-Weiss susceptibility of the Mn LMMs. The fitted effective PM moment of $5.0(2)\mu_B$ is in rough agreement with the high-spin configuration of $Mn^{2+}$ ($S$ = 5/2), while a small and positive value of the Curie-Weiss temperature ($\Theta_{CW}$ = 3(3) K) strongly depends on the fitted $\chi_0$ contribution. The $M(H)$ curve acquired below $T_N$ for $H \perp (0001)$ shows an indicative spin-flop transition at $\mu_0 H_{SF} \approx 3.5$ T (Fig. 2d), which is in line with an out-of-plane easy axis of the staggered magnetization.

We further performed X-ray magnetic circular dichroism (XMCD) experiments at the Mn $L_{2,3}$ absorption edge (Fig. 2e; D samples). The data were acquired in total electron yield (TEY) mode with a probing depth of typically only a few nm. The XMCD signal obtained at an external field of 6 T and in normal light incidence verifies a magnetic polarization of the Mn ions. After removing the external field ($H = 0$) the signal collapses as expected for an AFM ordering. Yet, a small residual signal is observed in remanence, indicating a finite net out-of-plane polarization within the probed volume of the sample. This residual signal appears to be inconsistent with an AFM intralayer coupling where the orientation of the moments within a Mn layer alternates on the atomic scale. For FM intralayer coupling, however, the first SL, which is preferentially probed in the TEY mode, is expected to be composed of mesoscopic domains with the magnetization pointing in or out of the surface plane. We attribute the residual XMCD signal in remanence to a preferential sampling of one domain type with the micron-sized synchrotron beam spot. This supports our first-principles calculations predicting an FM ordering within individual SLs. Performing the same experiment in grazing light incidence, i.e. with sensitivity to in-plane magnetization, we observe no finite polarization in remanence.

Altogether, the experimental scope of evidence (Fig. 2) allows identifying $MnBi_2Te_4$ as an interlayer antiferromagnet, in which FM Mn layers are coupled antiparallel to each other and the easy axis of staggered magnetization points perpendicular to the layers (Fig. 1d).

The surface electronic structure of $MnBi_2Te_4(0001)$ was studied using angle-resolved photoemission spectroscopy (ARPES). The intensity map measured near the Brillouin zone center at a temperature of 17 K is shown in Fig. 3a ($h\nu$ = 28 eV; B sample). One can clearly see two almost linearly dispersing bands forming a Dirac-cone-like structure with strongly reduced intensity at the crossing point. The energy distribution curves (EDCs) reveal an energy gap of about 70 meV at the $\bar{\Gamma}$-point that separates the upper and lower parts of the cone (Fig. 3b). A similar result was obtained for the D samples (Extended Data Fig. 4). These results agree with those of the $MnBi_2Te_4$(0001) surface bandstructure calculations (see Fig. 1f).

Next, we performed extensive ARPES measurements with different photon energies. At $h\nu$ = 9 eV (Fig. 3c), the features different from those seen at $h\nu$ = 28 eV show a pronounced spectral weight: namely, the intense electron- and hole-like bands coming to the $\bar{\Gamma}$-point at binding energies of about 0.17 eV and 0.4 eV, respectively. A comparison with the theoretically calculated bulk-projected bandstructure allows identifying these bands as the bulk conduction and valence bands, respectively. The analysis of the $\bar{\Gamma}$-point EDC shows that both valence and conduction bands can be fitted with two peaks (Extended Data Fig. 5) in agreement with the result of our calculations, showing two bulk bands with a weak $k_z$ dispersion both below and above the Fermi level (Fig. 1f). Based on the photoemission measurements, we estimate the bulk bandgap to be close to 200 meV, again, in agreement with the calculated values. The second derivative data in the inset to Fig. 3c provide further insight, apart from the bulk bands revealing also the gapped Dirac cone. Note that the Fermi level cuts the conduction band, indicating the *n*-doped character of the samples, which is consistent with the results of the Hall measurements (see Extended Data Fig. 3b). If we artificially drive $MnBi_2Te_4$ into a topologically-trivial phase by decreasing the SOC strength in a calculation, a strong discrepancy arises between the theory and experiment. Namely, in the trivial phase there are no surface states in the fundamental bulk bandgap (Extended Data Fig. 6). In contrast, the ARPES data acquired at different photon energies (Extended Data Fig. 7) unambiguously confirm the gapped Dirac cone to be a surface state, in agreement with the calculated (0001) surface bandstructure of the $MnBi_2Te_4$ AFMTI (Fig. 1f).

Thus, in virtue of the good agreement between the theory and experiment, we conclude that the fundamental bandgap of $MnBi_2Te_4$ is inverted. Our data also confirm the trigonal structure (sp. gr. $R\bar{3}m$) and interlayer AFM order, whereby the *S*-symmetry is confirmed, too. All this proves the $Z_2 = 1$ 3D AFMTI phase in $MnBi_2Te_4$ below its Néel temperature, which thus is the first experimentally identified system in the AFMTI class.

The ARPES measurements above $T_N$ were performed next. Remarkably, increasing the temperature does not lead to the DP gap closing at $MnBi_2Te_4$(0001), see Fig. 3d. Nevertheless, we observe a temperature dependence of a linear dichroism in the Dirac cone intensity that indicates an influence of the AFM order on the topological surface state (see Extended Data Fig. 8). Moreover, straight below $T_N$, the intensity of the Dirac cone starts to grow abruptly when the temperature decreases (Fig. 3e and Extended Data Fig. 9; laser ARPES with $h\nu$ = 6.4 eV), signaling the paramagnet-antiferromagnet transition and showing a clear response of the surface electronic structure to the AFM ordering. In Fig. 3f, the photoemission intensity map acquired at $T$ = 10.5 K is shown. A closer look at these data around the DP gap (Fig. 3g) and their comparison to the analogous spectrum taken at 35 K (Fig. 3h) reveals that significant changes occur across the Néel temperature. In particular, the shape of the $\bar{\Gamma}$-point EDC is modified: a pronounced intensity dip due to the DP gap at ~0.28 eV seen at 10.5 K (Fig. 3g) is absent at 35 K (Fig. 3h). Nevertheless the DP gap is present in Fig. 3h, as the EDC fit was only possible with two peaks, corresponding to the DP gap edges. Thus, while the antiferromagnet-paramagnet transition does not lead to the DP gap closing, the results in Figs. 3e-h reveal clearly the strong sensitivity of the topological surface state to the AFM ordering. The presence of a DP gap in the PM phase was previously reported for the surface states of magnetically-doped TIs[13-15]. In the case of the $Bi_{2-x}Mn_xSe_3(0001)$ surface, resonant scattering processes due to impurity in-gap states were suggested to be a possible reason of such behavior[15]. To check whether similar effects take place in our $MnBi_2Te_4$ samples, we performed resonant photoemission measurements at the Mn 3*p*–3*d* edge. The results, shown in Fig. 3i, reveal no resonant features and, hence, no Mn-3*d*-DOS around the DP gap, whereupon we discard such a mechanism of the DP gap opening in $MnBi_2Te_4$.

To get a deeper insight into the electronic properties of $MnBi_2Te_4$ both below and above $T_N$, we performed electron spin resonance (ESR) measurements (Fig. 4a; D samples). In $MnBi_2Te_4$, the ESR linewidth $\Delta H(T)$ shows a quite surprising anisotropic behavior above ~50 K. Namely, the so-called Korringa slope $b = d[\Delta H(T)]/dT$ of the linear $\Delta H(T)$ dependence for $H \perp (0001)$ is three times larger than that for $H \parallel (0001)$ (Fig. 4a), suggesting spatially anisotropic fluctuations of the LMMs due to their coupling to conduction electrons[16,17]. The fact that in $MnBi_2Te_4$ $b_\perp \gg b_\parallel$ indicates a relatively short transversal relaxation time $\tau_2$ of the LMMs for the out-of-plane static field geometry and, thus, fast fluctuations of the Mn spins in the (0001) plane. In contrast, in the case of the in-plane field geometry, where the $\tau_2$ relaxation time characterizes Mn spin fluctuations perpendicular to the (0001) plane, the fluctuations appear to be significantly slower, as reflected in the smaller $b_\parallel$. Consequently, on a timescale of the photoexcitation process in ARPES (~$10^{-15}$ seconds), which is much shorter than the Mn spins relaxation time (~$10^{-10}$ seconds), the probability to find the Mn spin oriented perpendicular to (0001) may be larger than for other directions, which could effectively generate an instantaneous out-of-plane field acting on the topological surface state electrons even in the absence of long-range order. Indeed, our room temperature spin-ARPES measurements reveal the out-of-plane spin components both at the $\bar{\Gamma}$-point (Fig. 4b) and at finite $k_\parallel$ (see Extended Data Fig. 10). Moreover, the calculated paramagnons indicate a presence of FM correlations within individual SL blocks above $T_N$ (right inset to Fig. 4a). At that, the FM-correlated moments tend to point perpendicular to the (0001) plane, which is consistent with the spin-ARPES observations, revealing out-of-plane spin polarization near the Fermi level. We thus conclude that the instantaneous out-of-plane spin polarization observed in our experiments could be responsible for the persistence of the DP gap observed by ARPES in the PM state of $MnBi_2Te_4$.

Our experimental and theoretical results establish $MnBi_2Te_4$ as the first observed AFMTI. Although in our experiment $MnBi_2Te_4$ is *n*-doped (which is typical for the Bi-based bulk

TI crystals), it is *n*-doping that allows measuring both the topological surface state and the bulk energy gap with ARPES, which probes only the occupied states. A common strategy in synthesizing truly insulating TI crystals is Sb-doping of the Bi sublattice of tetradymite-like compounds[2,18], which is expected to work for $MnBi_2Te_4$ as well. Note that such a tuning of composition is not supposed to affect its interlayer AFM ordering[19]. On the other hand, recent progress in the molecular beam epitaxy growth of the TI films[20] rises hope that a nearly charge neutral $MnBi_2Te_4$ can be fabricated.

All this would constitute an important step towards many novel applications. Indeed, according to Ref. [6], an AFMTI with the type of antiferromagnetic order established here for $MnBi_2Te_4$ represents ideal platform for observing the half-integer quantum Hall effect ($\sigma_{xy} = e^2/(2h)$), which may facilitate experimental confirmation of the $\theta = \pi$ quantized magnetoelectric coupling. A material showing this effect is known as an axion insulator, which up to now was being sought for in magnetically-doped sandwich-like TI heterostructures[21,22]. Unfortunately, the latter were found to show superparamagnetic behavior[4,23] – a drawback that $MnBi_2Te_4$ does not suffer from, which makes it a promising intrinsic axion insulator candidate. On the other hand, the FM SL blocks of $MnBi_2Te_4$ can be utilized for the fabrication of topologically-nontrivial heterostructures[24,25], which are promising for the quantum anomalous Hall effect[2]. Beyond topotronics, another direction of further studies of $MnBi_2Te_4$ and related materials[19,26-29] lies within the rapidly growing field of van der Waals magnets[30,31]. Strongly thickness-dependent properties, expected for the van der Waals compounds in the 2D limit, combined with the magnetic degrees of freedom and strong SOC of $MnBi_2Te_4$, make it an interesting candidate to couple the emerging fields of antiferromagnetic spintronics[32,33] and layered van der Waals materials[30,31].

We have become aware of recent theoretical[34,35] and experimental[36-42] studies on $MnBi_2Te_4$ that confirm our results. In particular, the intrinsic axion insulator state has just been realized in the AFM (i.e. below the Néel temperature) $MnBi_2Te_4$ thin flakes exfoliated from a bulk single crystal[41], providing an additional proof of the AFMTI state in $MnBi_2Te_4$ below $T_N$. Besides, the quantized Hall effect under external magnetic field has been achieved in such flakes[40-42]. Although the observation of this effect requires ferromagnetic state of $MnBi_2Te_4$, it would not be observed if $MnBi_2Te_4$ was topologically trivial[43].

**Acknowledgments**

M.M.O. and E.V.C. thank A. Arnau and J. I. Cerdá for stimulating discussions. We acknowledge the support by the Basque Departamento de Educacion, UPV/EHU (Grant No. IT-756-13), Spanish Ministerio de Economia y Competitividad (MINECO Grant No. FIS2016-75862-P), and Academic D.I. Mendeleev Fund Program of Tomsk State University (Project No. 8.1.01.2018). The supports by the Saint Petersburg State University grant for scientific investigations (Grant No. 15.61.202.2015), Russian Science Foundation (Grant No. 18-12-00062), Russian Foundation for Basic Research (Grant No. 18-52-06009), and the Science Development Foundation under the President of the Republic of Azerbaijan (Grant No. EİF-BGM-4-RFTF-1/2017-21/04/1-M-02) are also acknowledged. M.M.O. acknowledges support by the Diputación Foral de Gipuzkoa (project no. 2018-CIEN-000025-01). I.I.K. and A.M.S. acknowledge partial support from the CERIC-ERIC consortium for the stay at the Elettra synchrotron. The support by the German Research Foundation (DFG) is acknowledged by A.U.B.W., A.I., and B.B. within the Collaborative Research Center 1143, by A.Z., A.E., and A.I. within of the Special Priority Program 1666 Topological Insulators, by H.B. and F.R. within the Collaborative Research Center 1170, and by A.Z. and A.I. within the ERANET-Chemistry Program (RU 776/15-1). H.B., A.U.B.W., A.A., V.K., B.B., F.R. and A.I. acknowledge financial support from the DFG through Würzburg-Dresden Cluster of Excellence on Complexity and Topology in Quantum Matter - ct.qmat (EXC 2147, project-id 39085490). A.E. acknowledges support by the OeAD grants No. HR 07/2018 and PL 03/2018. This work was also supported by the Fundamental Research Program of the State Academies of Sciences, line of research III.23. A.K. was financially supported by KAKENHI number 17H06138 and 18H03683. I.P.R. acknowledges support by the Ministry of Education and Science of the Russian Federation within the framework of the governmental program Megagrants (state task no. 3.8895.2017/P220). The calculations were performed in Donostia International Physics Center, in the Research park of St. Petersburg State University Computing Center (http://cc.spbu.ru), and in Tomsk State University.

**Author Contributions**

The bandstructure calculations were performed by M.M.O., M.B.R., S.V.E, and A.Yu.V. The exchange coupling constants calculations were performed by M.B.R., Yu.M.K, M.M.O., and A.E. The paramagnons calculations were performed by A.E. The magnetic anisotropy studies were performed by M.M.O. The Monte Carlo simulations were performed by M.H. The topological invariant calculations were done by S.V.E. Tight-binding calculations were performed by I.P.R. and V.M.K. Crystals were grown by A.Z., A.I., Z.S.A., and M.B.B. X-ray diffraction measurements and structure determination were performed by A.Z., and I.R.A. The resistivity and Hall measurements, as well as contact preparation were done by N.T.M., N.A.A., and V.N.Z. XMCD and resonant photoemission experiments were performed by R.C.V., T.R.F.P., C.H.M., S.M., K.K., S.S., and H.B. Magnetization experiments and their analysis were mainly performed by S.G., B.B., and A.U.B.W. with contributions by A.V.K. ARPES measurements were done by I.I.K., D.E., A.M.S., E.F.S., S.K., A.K., L.P., G.D.S., H.B., and M.Ü. The analysis of the ARPES data was done by I.I.K., D.E., H.B., and M.Ü. Spin-ARPES measurements were performed by I.I.K, D.E., A.M.S., F.F., and J.S.-B. ESR measurements were done by A.A. and V.K. The project was planned by M.M.O., A.I., H.B., A.M.S., N.T.M., F.R., P.M.E., and E.V.C. The supervision of the project was executed by E.V.C. All authors contributed to the discussion and manuscript editing. The paper was written by M.M.O. with contributions by A.I., H.B., V.K. and A.U.B.W.

## Reprints and permissions

## Competing financial interest

Authors declare no competing financial interests.

## Correspondence and requests

Correspondence and requests for materials should be addressed to mikhail.otrokov@gmail.com.

## Main figure legends

**Figure 1. Theoretical insight into the crystal, magnetic and electronic structure of $MnBi_2Te_4$. a**, Crystal structure of the trigonal $MnBi_2Te_4$ with yellow, blue and green balls showing Mn, Te and Bi atoms, respectively. The "nonmagnetic" rhombohedral primitive unit cell is shown by grey lines, the $T_{1/2}$ translation is shown in magenta; "vdW" stands for van der Waals. **b**, Calculated exchange constants $J_{0,j}$ (in meV/$\mu_B^2$) for the intralayer ($J^{\parallel}$, red circles) and interlayer ($J^{\perp}$, blue squares) pair interactions as a function of the Mn-Mn distance, $r_{0,j}$ (in Å). **c**, Schematic top view representation of magnetic interactions in the Mn layer (upper part) and between two neighboring Mn layers (lower part). **d**, Magnetic unit cell corresponding to the interlayer AFM state. **e**, Total DOS of bulk $MnBi_2Te_4$ calculated for the interlayer AFM state shown in **d** using the HSE06 exchange-correlation functional (dash-point black line) and the GGA+$U$ approach (solid color lines). In the latter case, the evolution of the DOS with the change of the SOC constant $\lambda$ is shown. **f**, Spin-resolved electronic structure of the $MnBi_2Te_4$(0001) surface. The size of color circles reflects the value and sign of the spin vector Cartesian projections, with red/blue colors corresponding to the positive/negative $s_x$ and $s_y$ components (perpendicular to $k_{\|}$), and gold/cyan to the out-of-plane components $+s_z$/$-s_z$. The green areas correspond to the bulk bandstructure projected onto the surface Brillouin zone. The left-hand inset of panel **f** shows the tight-binding calculated electronic bandstructure of the $S$-preserving $(10\bar{1}1)$ surface (see also Extended Data Fig. 2). The regions with a continuous spectrum correspond to the 3D bulk states projected onto the 2D Brillouin zone.

**Figure 2. $MnBi_2Te_4$ single crystals and their magnetic properties. a, b**, $MnBi_2Te_4$ single crystals: D sample (**a**, optical microscope image) and B sample (**b**, scanning electron microscope image). **c**, Magnetic susceptibility (left axis of ordinates) of phase-pure $MnBi_2Te_4$ as a function of temperature measured in an external magnetic field of $\mu_0 H$ = 1 T in zero-field-cooled (zfc) and field-cooled-warming (fcw) conditions along with the temperature-dependent reciprocal susceptibility (right axis of ordinates) for $H \perp$ (0001). The green line is a modified Curie-Weiss fit to the high-temperature data ($\chi_0$ = 0.0028(3) emu/mol Oe); for details see text. **d**, Field-dependent magnetization curves for the two directions measured at 2 and 300 K. **e**, X-ray magnetic circular dichroism measurements for $MnBi_2Te_4$ at the Mn $L_{2,3}$ edge. A sketch of the experiment is given in the inset. The external magnetic field is applied along the direction of light incidence. The two top panels show the sum (XAS signal) and the difference (XMCD signal) between the $I_R$ and $I_L$ X-ray absorption intensities measured with right and left circularly polarized light in normal incidence at $\mu_0 H$ = 6 T. The bottom panels show the XMCD signal in remanence, i.e. at $H = 0$ T, for normal and grazing light incidence, measured after switching off an external field of $\mu_0 H$ = 6 T along the respective directions. Insets show a magnification of the $L_3$ dichroism in remanence.

**Figure 3. Photoemission spectroscopy insight into the surface and bulk bandstructure of $MnBi_2Te_4$. a**, Dispersion of $MnBi_2Te_4$(0001) measured at 17 K with a photon energy of 28 eV and, **b**, EDC representation of the data shown in **a**. The red curve marks the EDC at the $\bar{\Gamma}$-point. **c**, Dispersion of $MnBi_2Te_4$(0001) measured at the same temperature with a photon energy of 9 eV (which is more bulk sensitive) with the corresponding second derivative, $d^2N(E)/dE^2$, shown in the inset; BVB and BCB stand for the bulk valence and conduction bands, respectively. **d**, ARPES image acquired at 300 K ($hv = 28$ eV). **e**, Temperature dependence of the Dirac surface state photoemission intensity calculated as a sum of the intensities of the lower and upper parts of the cone at the $\bar{\Gamma}$-point (see Methods section for further details). **f**, The same as **c**, but measured with laser photon energy of 6.4 eV, $T = 10.5$ K, and a different sample. A small dashed rectangle around the DP marks the region that is further magnified in **g**. **g**,**h** Zooms into the DP gap region of the ARPES maps taken at 10.5 and 35 K, respectively, with the fitted EDC spectra at the $\bar{\Gamma}$-point plotted on top. The raw data, resulting fitted curves, and their decompositions with Voigt peaks are shown by blue circles, black/white lines, and grey/red lines, respectively. Red (grey) lines indicate the peaks attributed to the gapped Dirac cone state (bulk bands). The results shown in **a**-**h** were acquired on B samples. **i**, Resonant valence band spectra of $MnBi_2Te_4$ taken at the Mn 3$p$–3$d$ absorption edge (D samples). On- and off-resonance spectra were obtained at $hv = 51$ eV and $hv$ = 47 eV, respectively. The difference between these spectra approximately reflects the Mn 3$d$ DOS showing a main peak near 3.8 eV and an additional feature near 1 eV. The energy range of the bulk energy gap is marked by a cyan stripe.

**Figure 4. Spin characterization of bulk and surface electronic structure of $MnBi_2Te_4$. a**, Typical ESR spectra (field derivative of the microwave absorption $dP(H)/dH$) of $Mn^{2+}$ in $MnBi_2Te_4$ at $\nu = 9.6$ GHz (left inset) and $T$-dependence of the width $\Delta H$ of the ESR signal (main panel) for two directions of the magnetic field. Dashed lines are linear fits to the $\Delta H(T)$ dependence yielding the Korringa slopes $b_{\perp,\parallel} = d[\Delta H(T)/dT]$. The error bars are defined from the fitting procedure of ESR line shapes. The larger error bars above ~100 K are due to the occurrence of a parasitic signal in the ESR spectrum overlapping with the main line. The measurements were done on D samples. See Methods section for more information about ESR measurements of $MnBi_2Te_4$. The right inset of panel **a** shows a magnon spectrum calculated *ab initio* for bulk $MnBi_2Te_4$ in the PM state (paramagnons). **b**, Room temperature spin-resolved ARPES spectra measured at the $\bar{\Gamma}$-point with respect to the out-of-plane spin quantization axis. The out-of-plane spin polarization is presented below the corresponding spin-up and spin-down spectra. The inset shows the spin-integrated ARPES spectrum taken along the $\bar{K} - \bar{\Gamma} - \bar{K}$ direction. Yellow and cyan curves show the location of the gapped Dirac cone. The data were obtained for B sample at photon energy of 6 eV. See Extended Data Fig. 10 for more spin-resolved ARPES data.

## METHODS

### Electronic structure and total-energy calculations

Electronic structure calculations were carried out within the density functional theory using the projector augmented-wave (PAW) method[44] as implemented in the VASP code[45,46]. The exchange-correlation energy was treated using the generalized gradient approximation[47]. The Hamiltonian contained scalar relativistic corrections and the SOC was taken into account by the second variation method[48]. In order to describe the van der Waals interactions we made use of the DFT-D2[49] and the DFT-D3[50,51] approaches, which gave similar results. The energy cutoff for the plane-wave expansion was set to 270 eV. All structural optimizations were performed using a conjugate-gradient algorithm and a force tolerance criterion for convergence of 0.01 eV/Å. SOC was always included when performing relaxations.

The Mn 3$d$-states were treated employing the GGA+$U$ approach[52] within the Dudarev scheme[53]. The $U_{eff} = U - J$ value for the Mn 3$d$-states was chosen to be equal to 5.34 eV, as in previous works[24-26]. Besides, using this $U_{eff}$ we found a good agreement with the HSE06 functional[54-56] in the fundamental bandgap and binding energy of the Mn 3$d$-states of bulk $MnBi_2Te_4$. Also, we find a very good agreement of the calculated bandgap for the single $MnBi_2Te_4$ SL block (0.32 eV[43]) with the measured one (0.35 eV[36]). Moreover, the indirect character of the gap is correctly reproduced as well. Note that the GGA itself describes the bandgap of a single SL $MnBi_2Te_4$ film unsatisfactorily both in terms of the character (it yields direct gap) and size (43 meV). Thus, the use of GGA+$U$ improves the description of the *p–d* hybridization in the system. Further extensive testing was performed to check the stability of the results against $U_{eff}$ variation. Namely, the bulk crystal structure was fully optimized for different $U_{eff}$ values (3, 4, and 5.34 eV) and then the magnetic ordering was studied. It was found that the AFM ground state does not change upon such variations of $U_{\mathrm{eff}}$ and the crystal structure. The $MnBi_2Te_4$ magnetic anisotropy was found to be stable against these variations as well.

To model the interlayer AFM structure in $MnBi_2Te_4$, we used a rhombohedral cell with 14 atoms. These calculations were performed with the 3D Brillouin zone sampled by a 9×9×9 *k*-point grid.

The magnetic anisotropy energy, $E_a = E_{diff} + E_d$, was calculated taking into account both the total-energy differences of various magnetization directions, $E_{diff} = E_{in\text{-}plane} - E_{out\text{-}of\text{-}plane}$, and the energy of the classical dipole-dipole interaction. To calculate $E_{diff}$, the energies for three inequivalent magnetization directions [Cartesian $x$, $y$ (in-plane) and $z$ (out-of-plane)] were calculated and $E_{diff}$ was determined as the difference $E_{in\text{-}plane} - E_z$, where $E_{in\text{-}plane}$ is the energy of the most energetically favorable in-plane direction of the magnetization. A *k*-mesh of 2197 points was chosen, and the total energies were calculated self-consistently for all considered directions. The calculations of $E_{diff}$ were done for the interlayer AFM state in the cell containing 14 atoms (Fig. 1d). The energy convergence criterion was set to $10^{-7}$ eV providing a well-converged $E_{diff}$ (up to a few tenth of meV) and excluding "accidental" convergence. A cutoff radius of at least 20 µm was used to calculate $E_d$.

The $Z_2$ invariant for the 3D AFMTI was calculated according to Ref. [57]. When spatial inversion symmetry is present in the system, as it is in the $MnBi_2Te_4$ case, the following $Z_2$ invariant $\zeta_0$ can be defined:

$$(-1)^{\zeta_0} = \prod_{\boldsymbol{k}_{inv} \in B-TRIM, n \in occ/2} \zeta_n(\boldsymbol{k}_{inv}),$$

where $\zeta_n$ is the parity of the $n$th occupied band. See Ref. [57] for further details.

The $MnBi_2Te_4$ semi-infinite surface was simulated within a model of repeating films separated by a vacuum gap of a minimum of 10 Å. A 56-atomic-layers-thick slab was used, which corresponds to 8 SLs. The interlayer distances were optimized for the utmost SL block of each surface. Both the structural optimizations and static electronic structure calculations were performed using a *k*-point grid of 11×11×1 in the two-dimensional Brillouin zone.

**Exchange coupling constants calculations**

For the equilibrium structures obtained with VASP, we calculated the Heisenberg exchange coupling constants $J_{0,j}$ also from first principles, this time using the full-potential linearized augmented plane waves (FLAPW) formalism[58] as implemented in Fleur[59]. We took the GGA+$U$ approach[60,61] under the fully localized limit[62]. For the self-consistent FLAPW basis set in the $MnBi_2Te_4$ compound we chose a dense 22×22×22 Monkhorst-Pack *k*-point sampling of the first Brillouin zone and a cutoff of 3.4 Hartree. The density and potential expansions were cut at 10.4 Hartree. Locally, muffin tin sphere radii values of 2.74 a.u. for Mn and 2.81 a.u. for Bi and Te atoms were used, and the partial wave functions were expanded up to cutoffs of $l = 8$. Mn, Bi, and Te contribute $4s3d$, $5s5p$, and $6s6p$ valence electrons, respectively. We verified that these settings reproduce with great accuracy the bandstructures obtained with VASP, both with and without SOC effects. The FM vs. AFM ordering energy differences are also in agreement.

The $J_{0,j}$ constants were extracted by Fourier inversion of the magnon energy dispersion for the $MnBi_2Te_4$ primitive cell[59,63,64] neglecting SOC. These dispersion energies, calculated in the force theorem approach, correspond to a constrained set of non-collinear spin configurations characterized by the magnon $q$-vectors of a 13×13×13 grid[59,65]. The reference self-consistent electron wave functions were obtained with a 12×12×12 $k$-grid. These grids provided magnon energies converged below 0.1 meV and allowed to add up to 150 neighboring atoms to the Fourier analysis, ensuring accurate $J_{0,j}$ values.

**Paramagnons calculations**

Paramagnetic fluctuations were calculated within a first-principles approach based on the coherent potential approximation[66-69]. The disordered local moment (DLM) method was used to take ensemble averages over the orientational configurations of the local moments[70]. Paramagnons are calculated considering the response of the DLM paramagnetic state to the application of an external, site-dependent, magnetic field[71].

***Ab-initio*-based tight-binding calculations**

*Ab-initio*-based tight-binding calculations were performed using the VASP package with the WANNIER90 interface[72,73]. The wannier basis chosen consisted of six spinor $p$-type orbitals $|p_x^{\uparrow}\rangle$, $|p_y^{\uparrow}\rangle$, $|p_z^{\uparrow}\rangle$, $|p_x^{\downarrow}\rangle$, $|p_y^{\downarrow}\rangle$, $|p_z^{\downarrow}\rangle$ of Bi and Te. The surface electronic band structure was calculated within the semi-infinite medium Green function approach[74,75].

**Monte Carlo simulations**

The Monte Carlo (MC) simulations were based on a classical Heisenberg Hamiltonian including a magnetic anisotropy energy $E_a$

$$H = -\frac{1}{2}\sum_{ij} J_{i,j}\, \boldsymbol{e}_i \cdot \boldsymbol{e}_j + \sum_{i} E_a\, (\boldsymbol{e}_i^z)^2,$$

where the magnetic moments at site $i$ and $j$ are described by unit vectors $\boldsymbol{e}_i$ and $\boldsymbol{e}_j$, respectively, and the magnetic coupling constants $J_{i,j}$ are determined by *ab initio* calculations as described above. The term of the magnetic anisotropy takes into account only the $z$ component of the magnetic moment. We created for the simulation of the bulk system a cluster with periodic boundary conditions. Therein, the cluster size was varied with the primitive unit cell repeated $N$ times in all directions with $N = 12, 14, ..., 20$. This variation of system size allowed us to avoid finite size effects. We have tentatively started the calculation at a temperature of 60 K and reduced $T$ stepwise until 0.001 K. At each temperature step, the thermal equilibrium was reached after 40 000 MC steps (chosen after comparing results for 20 000 to 100 000 MC steps for all systems). The same number of MC steps was then used to derive the observables after reaching thermal equilibrium. One MC step represents the creation of a new random spin direction, which is either accepted or not depending on the energy difference and the current temperature. The critical temperature was finally obtained from the peak in the magnetic susceptibility, which represented the clearest identification criterion.

## Crystal growth

*Dresden samples (D samples)*

High-quality bulk single crystals of $MnBi_2Te_4$ were grown from the melt by slow cooling of a 1:1 mixture of the binaries $Bi_2Te_3$ and α-MnTe. The binaries were synthesized by mechanical pre-activation and annealing of stoichiometric mixtures of the elements. Crystal size and quality were controlled via different cooling rates within a narrow temperature interval at around 600 °C and varying annealing times. Further details of crystal-growth optimization are reported elsewhere[76]. Single-crystal X-ray diffraction was measured on a four-circle Kappa APEX II CCD diffractometer (Bruker) with a graphite(002)-monochromator and a CCD-detector at $T$ = 296(2) K. Mo-Kα radiation ($\lambda$ = 71.073 pm) was used. A numerical absorption correction based on an optimized crystal description[77] was applied, and the initial structure solution was performed in JANA2006[78]. The structure was refined in SHELXL against $F_o^2$[79,80]. The crystal structure has been deposited in the joint Cambridge Crystallographic Data Centre/FIZ Karlsruhe under the depository number CSD-1867581. The structure refinement yields some degree of statistical cation disorder in the Mn and Bi positions in contrast to an earlier reported ordered model[11]. Mn/Bi antisite defects in two fully occupied cation positions do not, however, lead to a superstructure ordering or change of translational symmetry. Energy dispersive X-ray spectra (EDS) were collected using an Oxford Silicon Drift X-MaxN detector at an acceleration voltage of 20 kV and a 100 s accumulation time. The analysis was performed using the $P/B-ZAF$ standardless method (where $Z$ = atomic no. correction factor, $A$ = absorption correction factor, $F$ = fluorescence factor, and $P/B$ = peak to background model). EDS reproducibly results in the stoichiometric composition, ruling out a possibility of large compositional variations in our samples.

*Baku samples (B samples)*

The bulk ingot of the Baku sample was grown from the melt with a non-stoichiometric composition using the vertical Bridgman method. The pre-synthesized polycrystalline sample was evacuated in a conical-bottom quartz ampoule sealed under vacuum better than $10^{-4}$ Pa. In order to avoid the reaction of the manganese content of the sample with the silica container during the melting process, the inside wall of the ampoule was coated with graphite by thermal decomposition of acetone in an oxygen-poor environment. The ampoule was held in the "hot" zone (~680 °C) of a two-zone tube furnace of the $MnBi_2Te_4$ Bridgman crystal growth system for 8 h to achieve a complete homogenization of the melts. Then, it moved from the upper (hot) zone to the bottom (cold) zone with the required rate of 0.7 mm/h. Consequently, the bulk ingot with average dimensions of 3 cm in length and 0.8 cm in diameter was obtained. Further details are reported elsewhere[81]. The as-grown ingot was checked by X-ray diffraction (XRD) measurements and was found to consist of several single crystalline blocks. With the aid of XRD data high quality single crystalline pieces were isolated from different parts of the as-grown ingot for further measurements.

**Magnetic measurements**

The magnetic measurements as a function of temperature and magnetic field were performed on a stack of single crystals of $MnBi_2Te_4$ (D samples) using a SQUID (Superconducting Quantum Interference Device) VSM (Vibrating Sample Magnetometer) from Quantum Design. The temperature dependent magnetization measurements were acquired in external magnetic fields of 0.02 T and 1 T for both zero-field-cooled (zfc) and field-cooled-warming (fcw) conditions. A thorough background subtraction was performed for all the curves.

Part of the magnetic measurements were carried out in the resource center "Center for Diagnostics of Materials for Medicine, Pharmacology and Nanoelectronics" of the SPbU Science Park using a SQUID magnetometer with a helium cryostat manufactured by Quantum Design. The measurements were carried out in a pull mode in terms of temperature and magnetic field. The applied magnetic field was perpendicular to the (0001) sample surface.

**Resistivity measurements**

Resistivity measurements were done with a standard four-probe ac technique using a low-frequency ($f \sim 20$ Hz) lock-in amplifier. Contacts were attached with conducting graphite paste. The measurements were carried out in a temperature-variable cryostat at different values of magnetic field up to 8 T, generated by a superconducting solenoid and directed along the normal to the (0001) sample surface.

Temperature- and field-dependent resistivity measurements were performed on B samples (Extended Data Fig. 3a). The metallic-like behavior characteristic of the presence of free carriers is observed at $H = 0$ as the resistivity $\rho$ increases with rising temperature. This is consistent with the results of the Hall effect measurements yielding the $n$-type conductivity of these samples (Extended Data Fig. 3b). A well-defined kink at 25.4 K indicates a magnetic transition in agreement with the magnetization studies and Monte Carlo simulations. In a series of measurements under an external field $H \perp (0001)$ the kink shifts to lower temperatures as the field increases from 1 T to 3 T. Above the critical field of ~3 T to 4 T, the $\rho(T)$ curve slope is much steeper below $T_N$, which could be related to the observed spin-flop in the $M(H)$ curve (Fig. 2d).

**ESR measurements**

ESR experiments were performed with a commercial X-band ESR-spectrometer (EMX from Bruker) operating at a microwave frequency of 9.6 GHz and providing magnetic fields up to 0.9 T. It is equipped with a He-gas flow cryostat (Oxford Instruments) and a goniometer allowing temperature and angular dependent measurements between 4 and 300 K.

The temperature dependence of the resonance field $H_{res}$ of the ESR signal of $Mn^{2+}$ ions in $MnBi_2Te_4$ is measured for the out-of-plane $H \perp (0001)$ and in-plane $H \parallel (0001)$ orientations of the static magnetic field $H$. At high temperatures $H_{res}$ is practically isotropic within the error bars and the corresponding spectroscopic $g$-factor is very close to the spin-only value $g_s = 2$ as expected for the $Mn^{2+}$ ($3d^5$; $S = 5/2$; $L = 0$) ion[82]. However, below $T \sim 50$ K $H_{res}$ becomes anisotropic and, in particular, the resonance line rapidly shifts to smaller fields for the out-of-plane geometry indicating an onset of the quasi-static short-range magnetic correlations developing in $MnBi_2Te_4$ well above the AFM phase transition at $T_N$. By further lowering the temperature, the linewidth experiences critical broadening due to the slowing down of the spin fluctuations in the vicinity of $T_N$[83] (Fig. 4a). Finally, upon entering the AFM ordered state the intensity of the ESR signal rapidly decreases, because of the shifting of the spectral weight to the AFM collective resonance modes typically occurring at much higher frequencies than the paramagnetic resonance[84] and, thus, out of the frequency range of the used ESR setup. Observation of quasi-static short-range magnetic correlations in the ESR experiment is consistent with the strong spin fluctuation-driven spin scattering above $T_N$ found in the magneto-transport study of $MnBi_2Te_4$ in Ref. [37].

**ARPES measurements**

The ARPES experiments were carried out at the BaDElPh beamline[85] of the Elettra synchrotron in Trieste (Italy) and BL-1 of the Hiroshima synchrotron radiation center (Japan) using $p$-polarization of the synchrotron radiation and laser[86,87]. The photoemission spectra were collected on freshly cleaved surfaces. The base pressure during the experiments was better than $1\times10^{-10}$ mbar. Part of the ARPES experiments were also carried out in the resource center "Physical methods of surface investigation" (PMSI) at the Research park of Saint Petersburg State University.

Additional $h\nu$-dependent experiments on D samples (the data are shown in Extended Data Fig. 7) were performed at the MAESTRO endstation of the Advanced Light Source (ALS).

**Dichroic ARPES measurements**

The linear-dichroism ARPES measurements on D samples (Extended Data Fig. 8) were performed at the MAESTRO endstation of the Advanced Light Source (ALS). This kind of measurement allow us to explore influence of the AFM state on the wave function of the topological surface state (TSS). With $p$-polarized light incident in the $xz$-plane the linear dichroism (LD) in the photoelectron intensity can be defined as[88-91]:

$$LD(k_x, k_y, E) = I(k_x, k_y, E) - I(-k_x, k_y, E).$$

The LD is the intensity asymmetry relative to the *yz*-plane. In the present experiment the *yz*-plane is a crystalline mirror plane. Therefore, in the absence of a magnetization, the LD is induced by the mirror-symmetry breaking of the light electric field vector $\boldsymbol{\mathcal{E}} = (\mathcal{E}_x, 0, \mathcal{E}_z)$. The intensities along $\pm k_x$ can be written as $I(\pm k_x) = |T_z \pm T_x|^2$, where $T_z$ and $T_x$ are the photoemission matrix elements of the electric field components $\mathcal{E}_z$ and $\mathcal{E}_x$ between the photoelectron final state $\Phi_f$ and the initial state $\psi_i$ at a given $k_x$[88]. One then finds $LD = 4\Re(T_z^* T_x)$, implying that a change of either matrix element will also change the LD. In particular, a change of the TSS wave function across $T_N$, will manifest in the LD because the TSS enters the matrix elements as the initial-state wave function $\psi_i$. This is precisely what we observe in Extended Data Fig. 8a,b,c: at temperatures above $T_N$ an LD is already present, but it gets considerably enhanced below $T_N$. This effect is understood as the effect of the AFM order on $\psi_i$. In the present case, as described in the main manuscript, $T_N$ marks the transition between the AFM state and a paramagnetic state with anisotropic fluctuations. Our LD-measurements provide strong evidence for an effect of the AFM order on the TSS wave function.

Our theoretical calculations provide additional insight into the origin of the LD. The crystal structure of the (0001) surface of $MnBi_2Te_4$ has 3 mirror planes along the $\bar{\Gamma} - \bar{M}$ directions of the 2D BZ. In the nonmagnetic case, the presence of these mirror planes dictates that the out-of-plane spin components are zero for the $\bar{\Gamma} - \bar{M}$ directions, along which the spins are locked exclusively within the surface plane. However, for the $\bar{\Gamma} - \bar{K}$ directions the $s_z$ components are allowed by symmetry and therefore can coexist with the in-plane spins. As one can see in Ref. [24], in the case of an isostructural compound $GeBi_2Te_4$ *ab initio* calculations indeed reveal the presence (absence) of the out-of-plane spins for the $\bar{\Gamma} - \bar{K}$ ($\bar{\Gamma} - \bar{M}$) directions. At $GeBi_2Te_4$(0001), the out-of-plane spin components are especially pronounced out of the fundamental bandgap energy region, where the TSS coincides with the bulk states in energy (but not in $k_\parallel$). By analogy, the appearance of the out-of-plane spin components along $\bar{\Gamma} - \bar{K}$ can be expected for the TSS of $MnBi_2Te_4$(0001). Such a spin texture is required to respect the time-reversal symmetry if the AFM exchange field inherent to the material is artificially set to zero. In this case, the $s_z$ sign will change upon changing $+k_\parallel$ (right branch) to $-k_\parallel$ (left branch). When the time-reversal symmetry is broken and each $MnBi_2Te_4$ SL is ferromagnetically ordered, the right and left branches of the TSS interact differently with the Zeeman field provided by the Mn layer of the topmost SL block (Extended Data Fig. 8d). This is similar to a so-called exchange+Rashba effect[92,93]. In such cases, a dispersion asymmetry is created (i.e. $E(+k_\parallel) \neq E(-k_\parallel)$), which is indeed seen for the $MnBi_2Te_4$ TSS (Extended Data Fig. 8d,e), resulting in a sizable asymmetry in the orbital composition between its right and left branches (Extended Data Fig. 8e,f). Note that such asymmetries are not found in our calculations for the $\bar{\Gamma} - \bar{M}$ directions because of the symmetry constraints on the presence of $s_z$ components in the time-reversal symmetry preserving case. In the time-reversal symmetry breaking case, the $s_z$ components appear along the $\bar{\Gamma} - \bar{M}$ directions, which also breaks the mirror symmetry at the $MnBi_2Te_4$(0001) surface.

**Temperature-dependent laser ARPES measurements**

To study the Dirac cone state response to the onset of the AFM order, we carried out the temperature-dependent high-resolution laser ARPES measurements for the $MnBi_2Te_4$(0001) surface. In Extended Data Fig. 9a, the raw EDC profiles at the $\bar{\Gamma}$-point at 10.5 and 35 K are presented (the same as in Figs. 3g,h of the manuscript), clearly showing the increase of the intensities of both the upper and lower parts of the Dirac cone state below the Néel temperature. Before studying the Dirac cone intensity variation systematically, we have carefully analyzed the intensity of the first two bulk conduction band states (those shown in the Extended Data Fig. 5c). No temperature dependence of the integral intensity of these bulk states was observed (Extended Data Fig. 9b) and therefore the EDC spectra require no specific temperature-dependent normalization. Thus, the intensity temperature dependence was obtained using the raw ARPES data. The results are shown in Extended Data Fig. 9c. Here, the $\bar{\Gamma}$-point EDC profiles, acquired with a step of ~0.9 K, are plotted versus temperature. Below ~24-25 K, the intensity enhancement in the $\Delta E$=[0.2;0.3 eV] energy window, containing the Dirac cone state at the $\bar{\Gamma}$-point, is clearly seen. Note that these temperature-dependent ARPES spectra were measured with the equal acquisition time and the constant laser photon flux for each temperature point in Extended Data Fig. 9c. In order to get deeper insight into this behavior of intensity, we integrate it within the energy window $\Delta E$ and plot as a function of temperature in Extended Data Fig. 9d (the same data are presented in Fig. 3e of the manuscript). It is seen that while the Dirac cone intensity is roughly constant above $T_N \approx 24$ K, below the Néel temperature it starts to grow and at 10 K gets enhanced by a factor of 1.4-1.5 as compared to its value at 35 K.

It should be noted that during the ARPES measurements the spectra acquisition time is rather limited, which is due to a finite "lifetime" of the clean surface of the sample even under the ultrahigh vacuum conditions. This lifetime is well comparable to the time needed to acquire several high quality ARPES maps (such as shown in Figs. 3g,h), that are suitable for a reliable comparative EDC spectral analysis (fitting). However, within the lifetime of the clean surface of the sample, such a long acquisition time per temperature point is unreachable during the systematic temperature-dependent measurements that in our case are made with more than 50 temperature points, i.e. $\Delta T \approx 0.9$ K and two sweep directions (10 K → 35 K → 10 K). Although the measured EDCs shown in Extended Data Fig. 9c are obviously meaningful, their spectral decomposition (fitting), similar to that shown in Extended Data Fig. 9a, is not unambiguous. Treated without any normalization, these raw data clearly show a strong intensity change across the Néel temperature.

As it can be seen from the EDC analysis (Extended Data Fig. 9a), the intensity increase observed is largely due to the enhancement of the gapped Dirac cone peaks intensities, while the signal between those peaks (i.e. within the DP gap) is practically the same for 10.5 and 35 K. Therefore, at 35 K the intensity inside the DP gap increases *relative* to those of the Dirac cone peaks, indicative of an enhancement of the spectral weight inside the gap. The temperature-dependent change of the line shape around the DP gap can be nicely demonstrated by the behavior of the second derivative ($d^2N(E)/dE^2$), which is highly sensitive to the curvature of the spectrum $N(E)$, see Extended Data Fig. 9e. Indeed, at low temperatures one can see a pronounced feature of the second derivative around the binding energy of 0.25 eV, signaling the presence of the DP gap, while above the Néel temperature this feature is significantly weaker. These changes clearly demonstrate the gapped state mitigation across the antiferromagnet-paramagnet transition.

**Spin-ARPES measurements**

Spin-resolved ARPES measurements were performed at the RGBL-2 endstation of BESSY-II in Berlin using a Scienta R4000 hemispherical analyzer and a photon energy of 6 eV. The photon beam was generated using the fourth harmonic of a homemade fs-laser system coupled to an ultrafast amplifier operating at 100 kHz repetition rate. The spin-resolved spectra were acquired with a 3D Mott-type spin detector operated at 26 kV. We used the vertical linearly polarized light incident on the sample under an angle of 45° with respect to the surface normal. The experimental geometry is given in Ref. [94]. The energy and angular resolutions of the spin-resolved measurements were 30 meV and 1.5° (corresponds to 0.02 $Å^{-1}$), respectively.

**ResPES measurements**

ResPES data were acquired at the HR-ARPES branch of the I05 beamline at the Diamond Light Source. The measurements were conducted at a base temperature of $T = 10$ K with a beam spot size and energy resolution of $A_{spot} \approx 50\times50$ $\mu m^2$ and $E \approx 20$ meV, respectively. The difference of on- and off-resonant spectra for the Mn $3p$–$3d$ transition corresponds directly to the Mn $3d$ density of states. A photon energy series was conducted in order to determine suitable transition energies. The corresponding angle integrated spectra of on-resonant ($hv = 51$ eV) and off-resonant ($hv = 47$ eV) conditions can be seen in Fig. 3i.

**XMCD measurements**

Surface sensitive XMCD measurements (TEY mode[95]) on D samples were performed at the HECTOR end-station of the BOREAS beamline at the ALBA synchrotron radiation facility[96]. The data were collected in total electron yield mode. The spot size and the resolving power of the supplied photon beam were $A_{spot} < 200\times200$ $\mu m^2$ and $E/\Delta E > 9000$, respectively. Measurements were performed at the Mn-$L_{2/3}$ edges at a temperature of 2 K, i.e. well below $T_N \approx 24$ K.

**Data availability**

The data that support the findings of this study are available from the corresponding author upon reasonable request. The crystal structure is available in the joint Cambridge Crystallographic Data Centre/FIZ Karlsruhe under the depository number CSD-1867581.

## Extended Data figure legends

**Extended Data Figure 1. Monte Carlo simulations for bulk $MnBi_2Te_4$**. The temperature dependent-magnetic susceptibility of bulk $MnBi_2Te_4$ calculated for various numbers of magnetic shells $j$ up to which the exchange coupling constants $J_{i,j}$ were considered in the classical Heisenberg Hamiltonian. Up to 10, 20, ... 70 shells were included in the calculation. The vertical dashed line shows the final Néel temperature of 25.4 K as estimated for the calculation including 70 shells. Note that only above 20 shells, the Monte Carlo simulations determined the antiferromagnetic ground state.

**Extended Data Figure 2. Electronic structure of the *S*-breaking and *S*-preserving surfaces of $MnBi_2Te_4$.** Surface electronic bandstructure of $MnBi_2Te_4$ calculated for the (0001), **a** (*S*-breaking), and $(10\bar{1}1)$, **b** (*S*-preserving), terminations using the *ab-initio*-based tight-binding approach. The regions with a continuous spectrum correspond to the 3D bulk states projected onto a 2D Brillouin zone.

**Extended Data Figure 3. Resistivity and Hall measurements of bulk $MnBi_2Te_4$.** **a**, Temperature- and field-dependent resistivity data. **b**, Hall voltage as a function of the applied magnetic field for $MnBi_2Te_4$ at 5 K. The Hall effect measurements unambiguously indicate the *n*-type conductivity of our $MnBi_2Te_4$ samples that show a negative Hall voltage for positive values of the applied magnetic field. The estimated electron concentration and Hall mobility are ~$2\times10^{19}$ $cm^{-3}$ and ~100 $cm^2/(V\cdot s)$, respectively. The measurements were performed on B samples.

**Extended Data Figure 4. D sample ARPES. a,** Dispersion of $MnBi_2Te_4(0001)$ measured at 12 K with a photon energy of 28 eV and, **b**, EDC representation of the data shown in **a**. The red curve marks the EDC at the $\bar{\Gamma}$-point. **c**, ARPES image acquired at 300 K ($h\nu$ = 28 eV). **d**, Dispersion of $MnBi_2Te_4(0001)$ measured at a temperature of 12 K with a photon energy of 9 eV (which is more bulk sensitive) and, **e**, the corresponding second derivative, $d^2N(E)/dE^2$. **f**, ARPES image acquired at 300 K ($h\nu$ = 9 eV). The measurements were performed on D samples.

**Extended Data Figure 5. Laser-based ARPES. a,** ARPES map of $MnBi_2Te_4$ taken at 10 K with a photon energy of 6.3 eV. In this case, the photoelectrons have a kinetic energy of about 1.5 eV and, subsequently, a large mean free path in the sample, corresponding to a high bulk sensitivity of this experiment. **b**, second derivative $d^2N(E)/dE^2$ of the data shown in **a**. **c**, fitting results for the EDC spectrum at the $\bar{\Gamma}$-point. The raw data, resulting fitting curve and its decomposition with Voigt peaks are shown by blue symbols, black solid line and grey dashed/red solid lines, respectively. Red (grey) lines indicate the peaks attributed to the gapped Dirac cone state (bulk bands). The measurements were performed on B samples.

**Extended Data Figure 6. Surface electronic structure of $MnBi_2Te_4$ in the artificial topologically-trivial phase.** Septuple-layer resolved (0001) surface electronic structure of $MnBi_2Te_4$ calculated for the SOC constant value $\lambda = 0.55\lambda_0$. The size of the color circles reflects the state localization in a particular SL block of the 8-SL-thick slab: **a**, 1st

(i.e. surface; red), **b**, 2nd (subsurface; blue), **c**, 3rd (bulk-like; green), and **d**, 4th (bulk-like; black). The grey areas correspond to the bulk bandstructure projected onto the surface Brillouin zone. It is seen that near the $\bar{\Gamma}$-point there are (i) no surface states in the bulk bandgap and (ii) no resonance states near the bandgap edges. The first quantum well states of both the valence and conduction bands are strongly localized in the inner parts of the slab.

**Extended Data Figure 7. Photon-energy-dependent ARPES data.** Photon-energy-dependent ARPES data measured near the Brillouin zone center along the $\bar{K}-\bar{\Gamma}-\bar{K}$ direction at a temperature of 18 K. Absence of $h\nu$-dependence confirms surface-state character of the upper cone. The measurements were performed on D samples.

**Extended Data Figure 8. Temperature-dependent linear dichroism in the Dirac cone photoemission intensity.** **a**, Dispersion of $MnBi_2Te_4$(0001) measured at 18 K with a photon energy of 21.5 eV and *p*-polarized light along the $\bar{K}-\bar{\Gamma}-\bar{K}$ direction. **b**, Momentum distribution curves representation of the data acquired at 18 K (blue) and 80 K (red). **c**, Linear dichroism ($I_{right} - I_{left}$), where $I_{right}$ and $I_{left}$ are the intensities of the right and left branches of the upper and lower cone corresponding to the positive and negative $k_{||}$, respectively. The measurements were performed on D samples. **d**, Upper part of the $MnBi_2Te_4$(0001) gapped Dirac cone as calculated *ab initio*. The size of color circles reflects the value and sign of the spin vector Cartesian projections, with red/blue colors corresponding to the positive/negative $s_y$ components (perpendicular to $k_{||}$), and gold/cyan to the out-of-plane components $+s_z$/$-s_z$. **e**, The same as in **d**, but with the size of the purple circles reflecting the weight of the $p_x$-orbitals of all Bi and Te atoms of the topmost SL block at each $k_{||}$. Note that in **d,e** the bulk-like bands of the slab are omitted. The magnetic moment of the topmost Mn layer points towards vacuum, while in Fig. 1f of the manuscript and Extended Data Fig. 6 it points in the opposite direction. **f**, The weight of the $s$, $p_x$, $p_y$, and $p_z$ orbitals of all Bi and Te atoms of the topmost SL block for the left (triangles) and right (squares) branches as a function of energy. See Methods section for more information on the dichroic ARPES measurements.

**Extended Data Figure 9. Temperature-dependent laser ARPES measurements. a,** ARPES EDC profiles taken at the $\bar{\Gamma}$-point of $MnBi_2Te_4$(0001) at 10.5 and 35 K. The raw data, resulting fitted curves, and their decompositions with Voigt peaks are shown by color symbols, black dashed lines, and color lines/gray symbols, respectively. Red/blue lines (circles/squares) indicate the peaks (EDCs) of the Dirac cone state at 35/10.5 K. The peaks of the bulk bands at 35/10.5 K are shown by gray circles/squares. **b**, Integrated intensity of the first two bulk conduction band states (those analyzed in detail in the Extended Data Fig. 5c) as a function of temperature. The inset shows the ARPES $MnBi_2Te_4$(0001) map measured with laser photon energy of 6.4 eV and $T = 10.5$ K (the same as in Fig. 3f). The green rectangle marks the region of the map where the first two bulk conduction band states are located. The average intensity in the shown temperature interval was set to 1. **c**, EDC profiles, $N(E)$, taken at the $\bar{\Gamma}$-point between 10 and 35 K with a temperature step $\Delta T$ of ~0.9 K and two sweep directions (10 K → 35 K → 10 K). Since the measurements upon heating and cooling reveal essentially the same behavior,

in panel **c** we show the data averaged over these two sets of the EDC profiles at each temperature point. Note that the data in panel **a** and temperature intensity dependencies in panels **b**,**c**,**d** were acquired from two different samples, showing slightly different binding energy of the DP gap centers (0.28 vs 0.25 eV, respectively). **d**, Intensity integrated within the energy window $\Delta E$ marked by the dashed black lines in **c.** The average intensity in the "plateau-like" region above ~24 K was set to 1. $\Delta E$ contains both lower and upper parts of the Dirac cone at the $\bar{\Gamma}$-point and corresponds to the energy interval where the contribution of the cone is dominant, while that of the bulk states is almost negligible. The vertical cyan line approximately shows the start of the intensity increase, which roughly corresponds to $T_N \approx 24$ K of $MnBi_2Te_4$. **e**, The second derivative, $d^2N(E)/dE^2$, of the EDC profiles on **c**, shown for better visualization of the DP gap behavior.

**Extended Data Figure 10. Spin-resolved ARPES data.** **a**, Spin-integrated ARPES spectrum taken at 6 eV photon energy along the $\bar{K} - \bar{\Gamma} - \bar{K}$ direction. Yellow and cyan curves show the location of the gapped topological surface state; **b**, Spin-resolved ARPES spectra taken at the $\bar{\Gamma}$-point with respect to the out-of-plane spin quantization axis. The out-of-plane spin polarization is shown below the corresponding spin-up and spin-down spectra; **c**,**d**, Measured out-of-plane and in-plane spin polarization at different momentum values, respectively. The in-plane spin polarization changes its sign with $k_{\|}$, as expected for the topological surface state. The change of the out-of-plane spin polarization sign at $k_{\|} = +0.1$ Å$^{-1}$ near the Fermi level in panel **c** is discussed in the Methods section dedicated to the dichroic ARPES measurements description. The data on **a**,**b** and **c**,**d** were acquired on B and D samples, respectively. The measurements were performed at $T = 300$ K.

# Figure 1

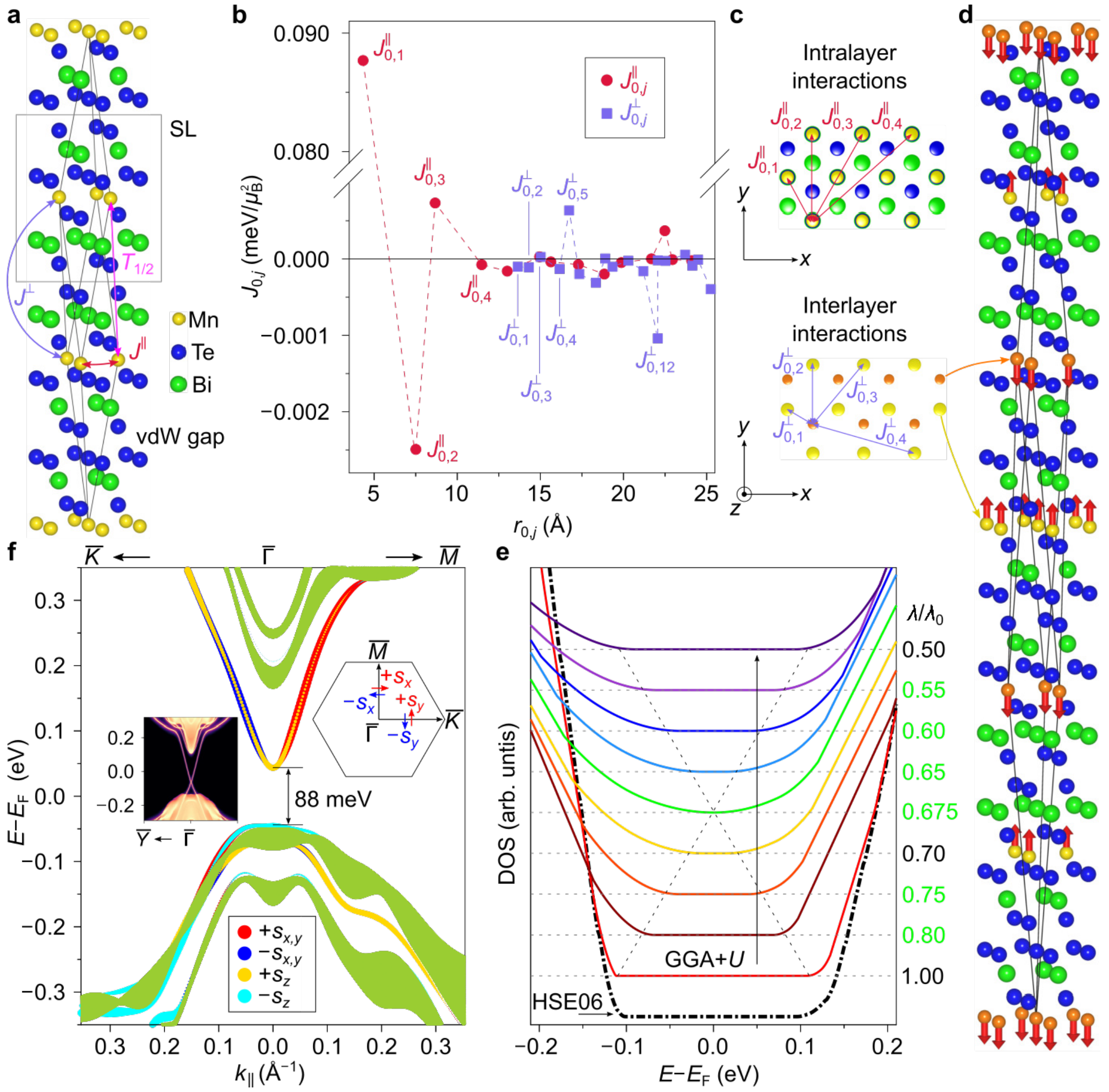

a
b
c
d
e
f
SL
vdW gap
Mn
Te
Bi
Intralayer interactions
Interlayer interactions
88 meV
GGA+U
HSE06
DOS (arb. untis)
λ/λ0
0.50
0.55
0.60
0.65
0.675
0.70
0.75
0.80
1.00

## Figure 2

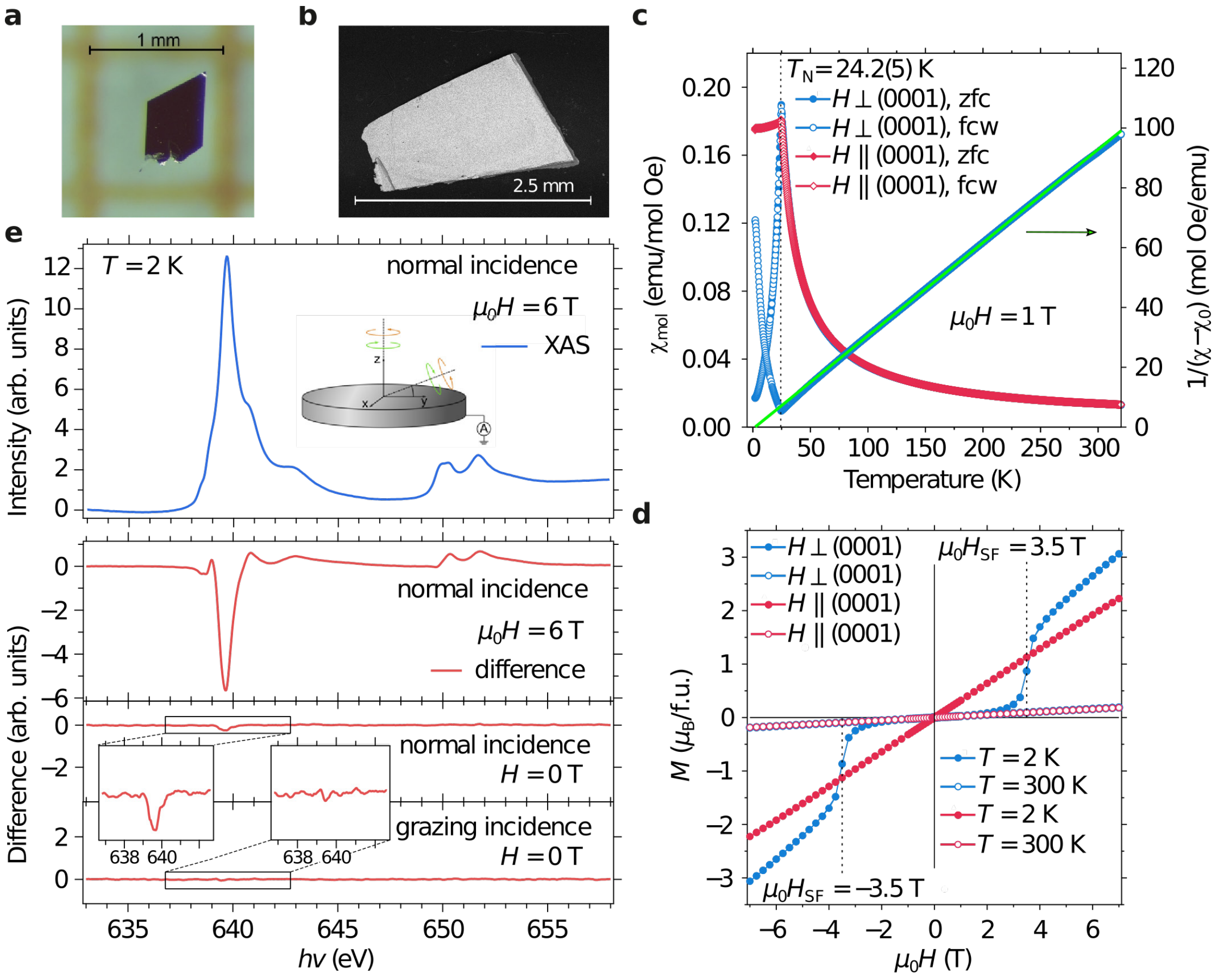

a
1 mm
b
2.5 mm
c
$T_N$=24.2(5) K
H ⊥ (0001), zfc
H ⊥ (0001), fcw
H || (0001), zfc
H || (0001), fcw
$\mu_0 H$ = 1 T
$\chi_{mol}$ (emu/mol Oe)
$1/(\chi - \chi_0)$ (mol Oe/emu)
Temperature (K)
e
T = 2 K
normal incidence
$\mu_0 H$ = 6 T
XAS
Intensity (arb. units)
normal incidence
$\mu_0 H$ = 6 T
difference
normal incidence
H = 0 T
grazing incidence
H = 0 T
Difference (arb. units)
hν (eV)
d
H ⊥ (0001)
H ⊥ (0001)
H || (0001)
H || (0001)
$\mu_0 H_{SF}$ = 3.5 T
T = 2 K
T = 300 K
T = 2 K
T = 300 K
$\mu_0 H_{SF}$ = −3.5 T
M ($\mu_B$/f.u.)
$\mu_0 H$ (T)

# Figure 3

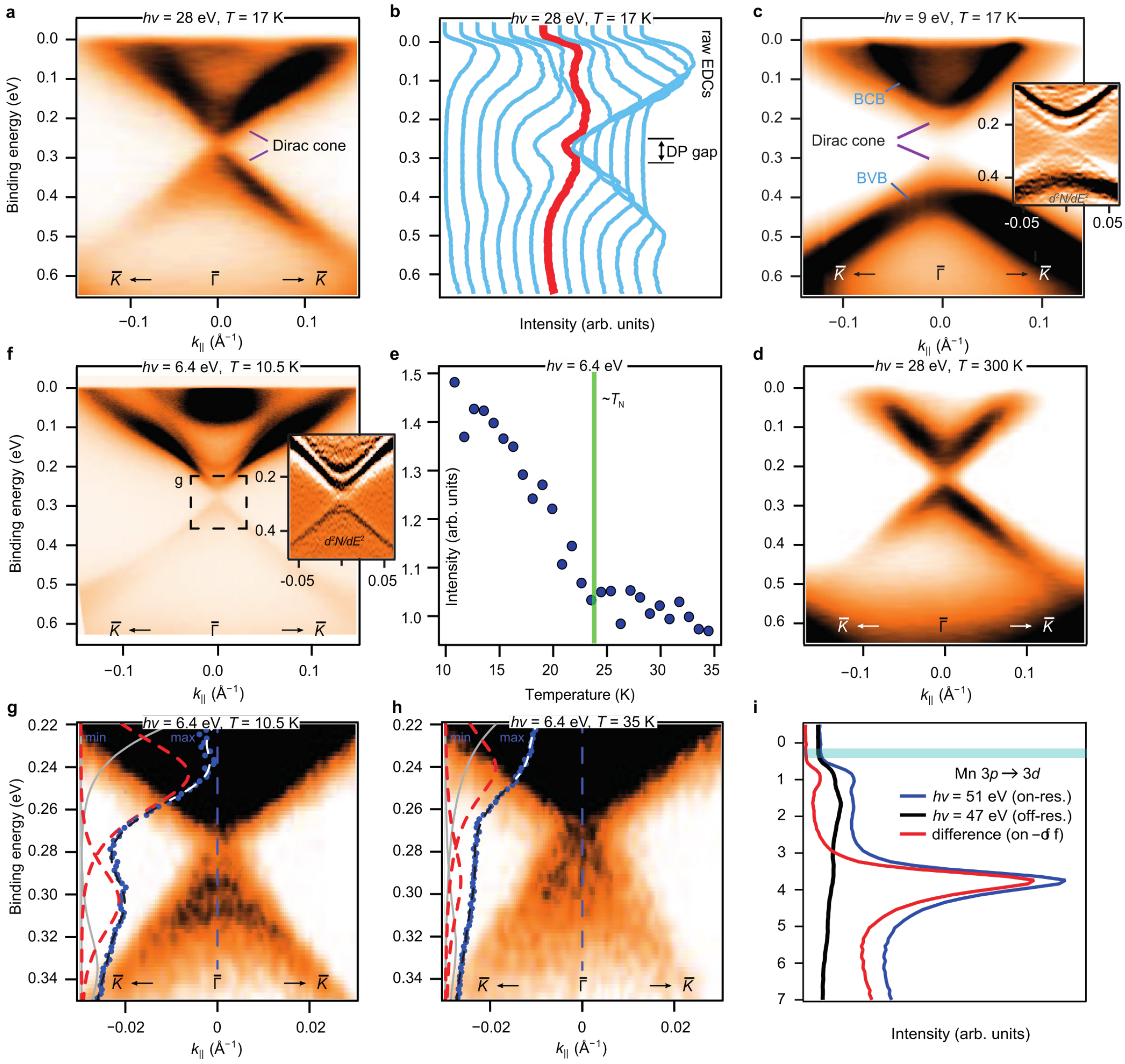

a
hν = 28 eV, T = 17 K
Binding energy (eV)
Dirac cone
k|| (Å⁻¹)
b
hν = 28 eV, T = 17 K
raw EDCs
DP gap
Intensity (arb. units)
c
hν = 9 eV, T = 17 K
BCB
Dirac cone
BVB
d²N/dE²
d
hν = 28 eV, T = 300 K
e
hν = 6.4 eV
~T_N
Intensity (arb. units)
Temperature (K)
f
hν = 6.4 eV, T = 10.5 K
g
d²N/dE²
g
hν = 6.4 eV, T = 10.5 K
min
max
h
hν = 6.4 eV, T = 35 K
i
Mn 3p → 3d
hν = 51 eV (on-res.)
hν = 47 eV (off-res.)
difference (on − off)
Intensity (arb. units)

**Figure 4**

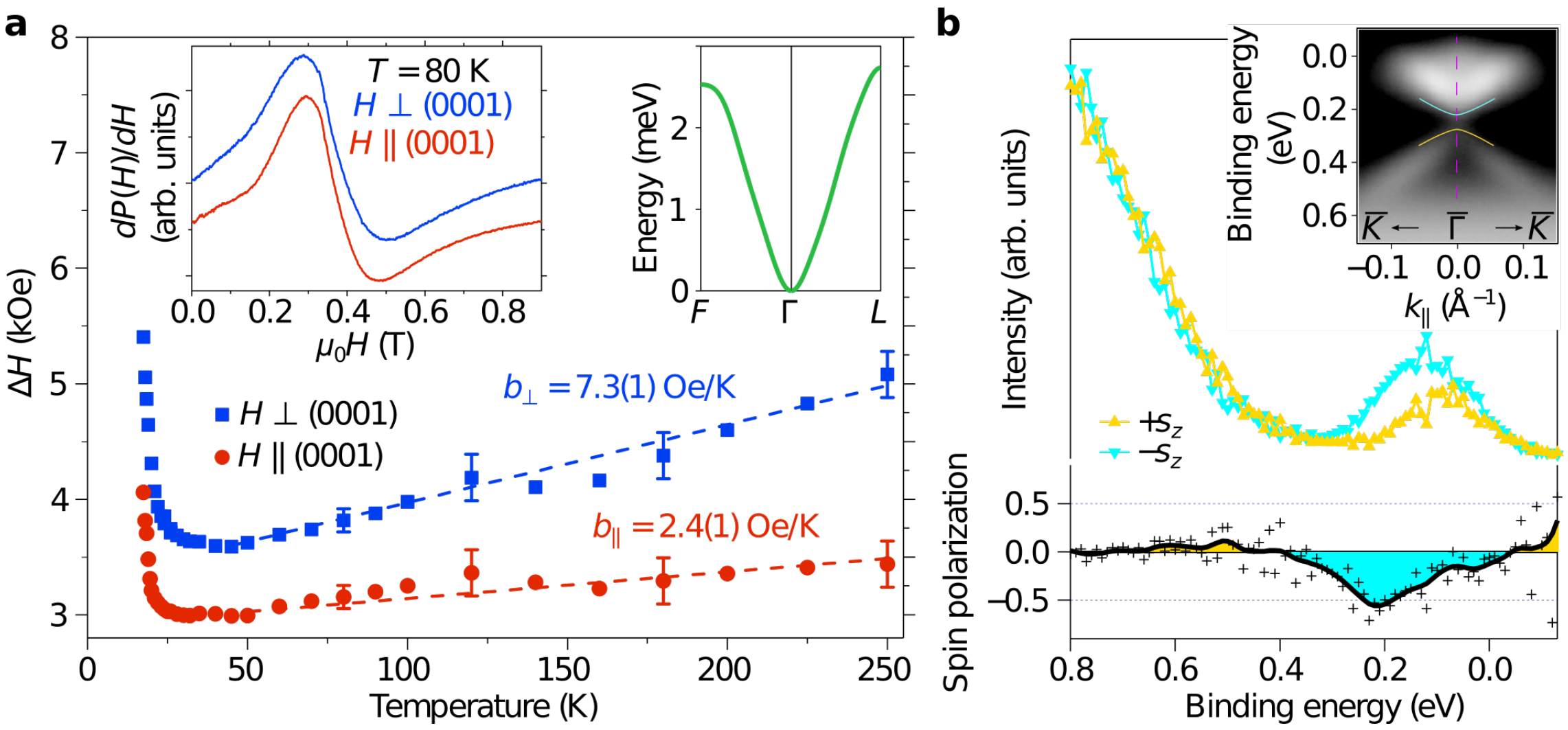

a
b
T = 80 K
H ⊥ (0001)
H || (0001)
dP(H)/dH (arb. units)
μ0H (T)
Energy (meV)
F
Γ
L
ΔH (kOe)
b⊥ = 7.3(1) Oe/K
b|| = 2.4(1) Oe/K
Temperature (K)
Intensity (arb. units)
+sz
−sz
Binding energy (eV)
k|| (Å−1)
Spin polarization
Binding energy (eV)

**Extended Data Fig. 1**

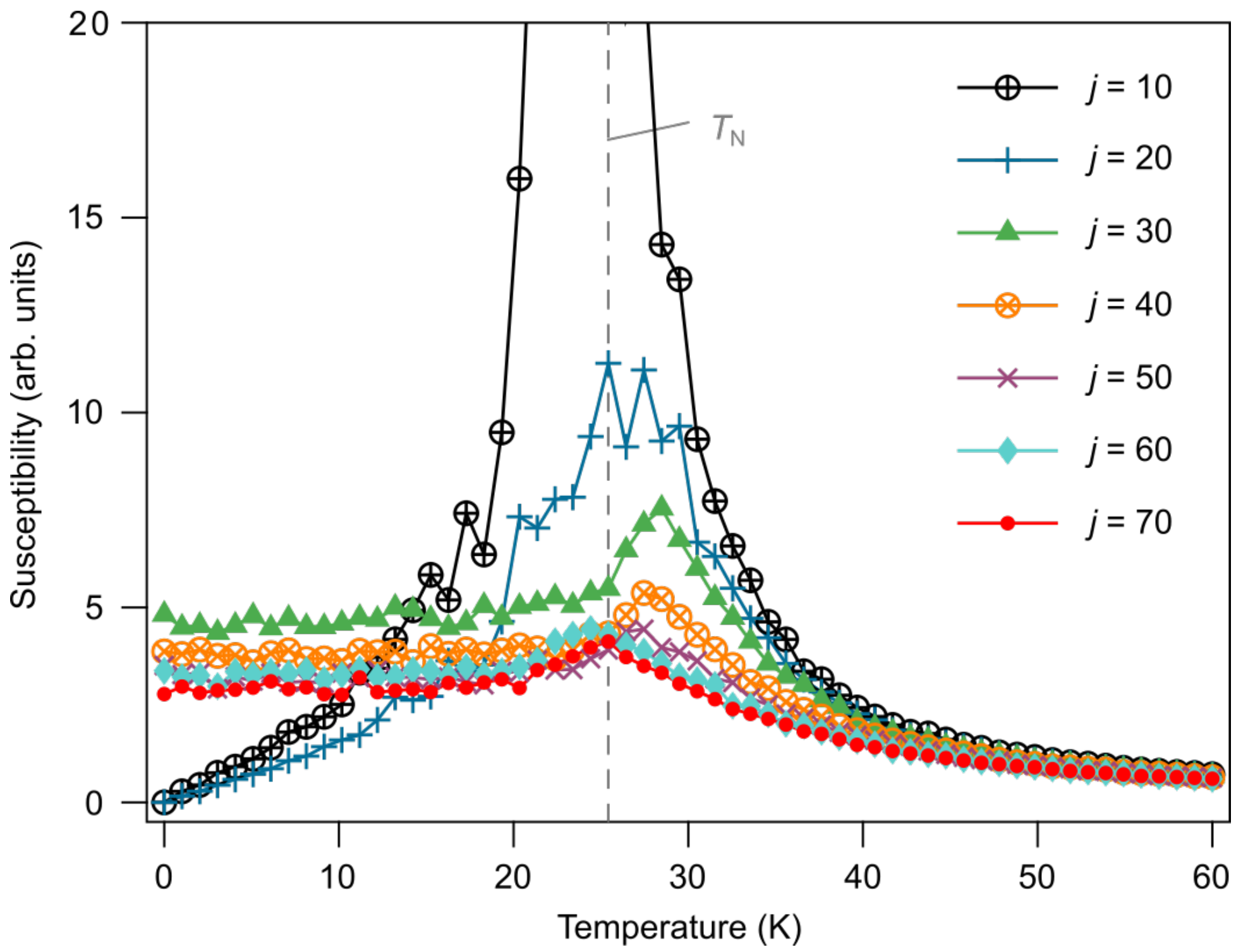

**Extended Data Fig. 2**

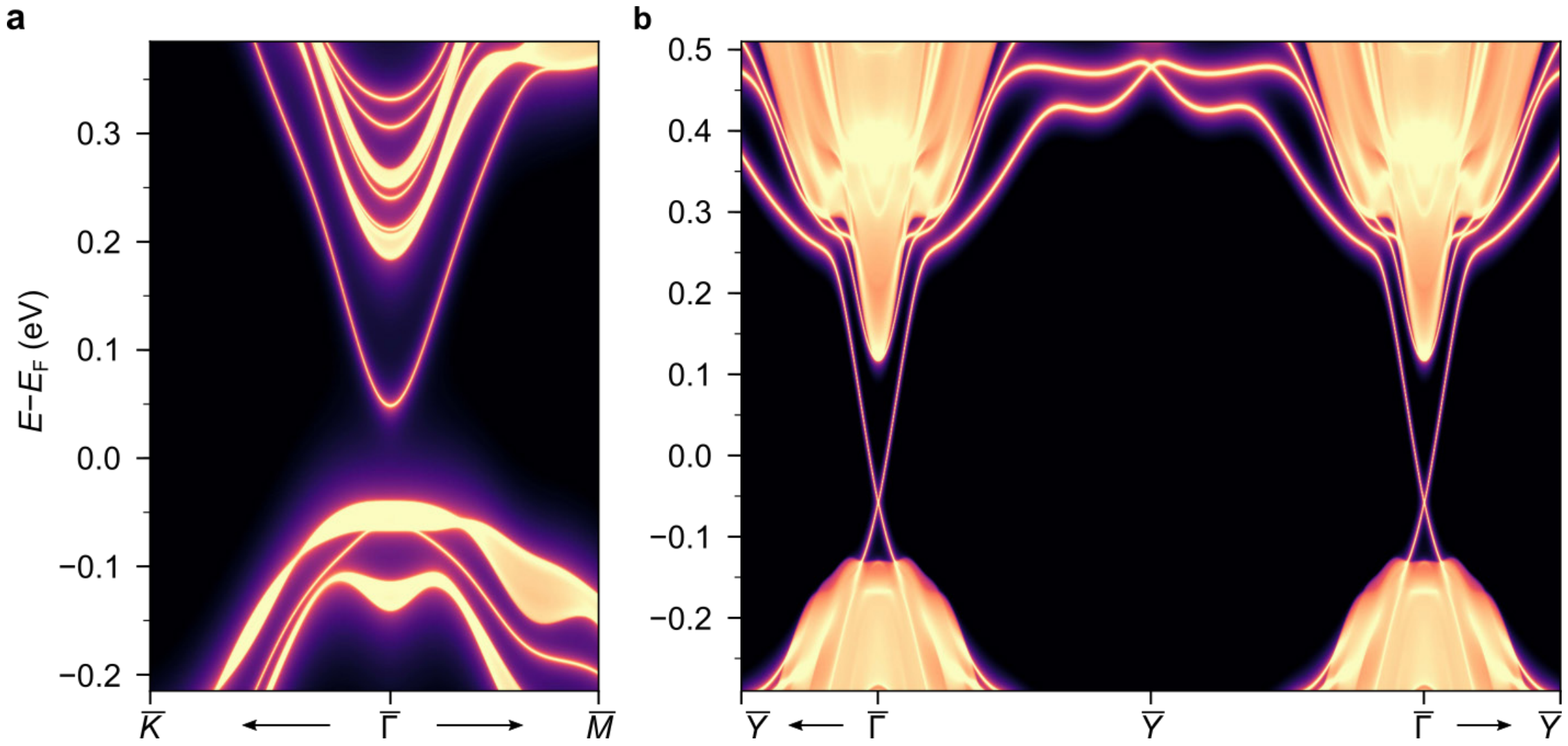

**Extended Data Fig. 3**

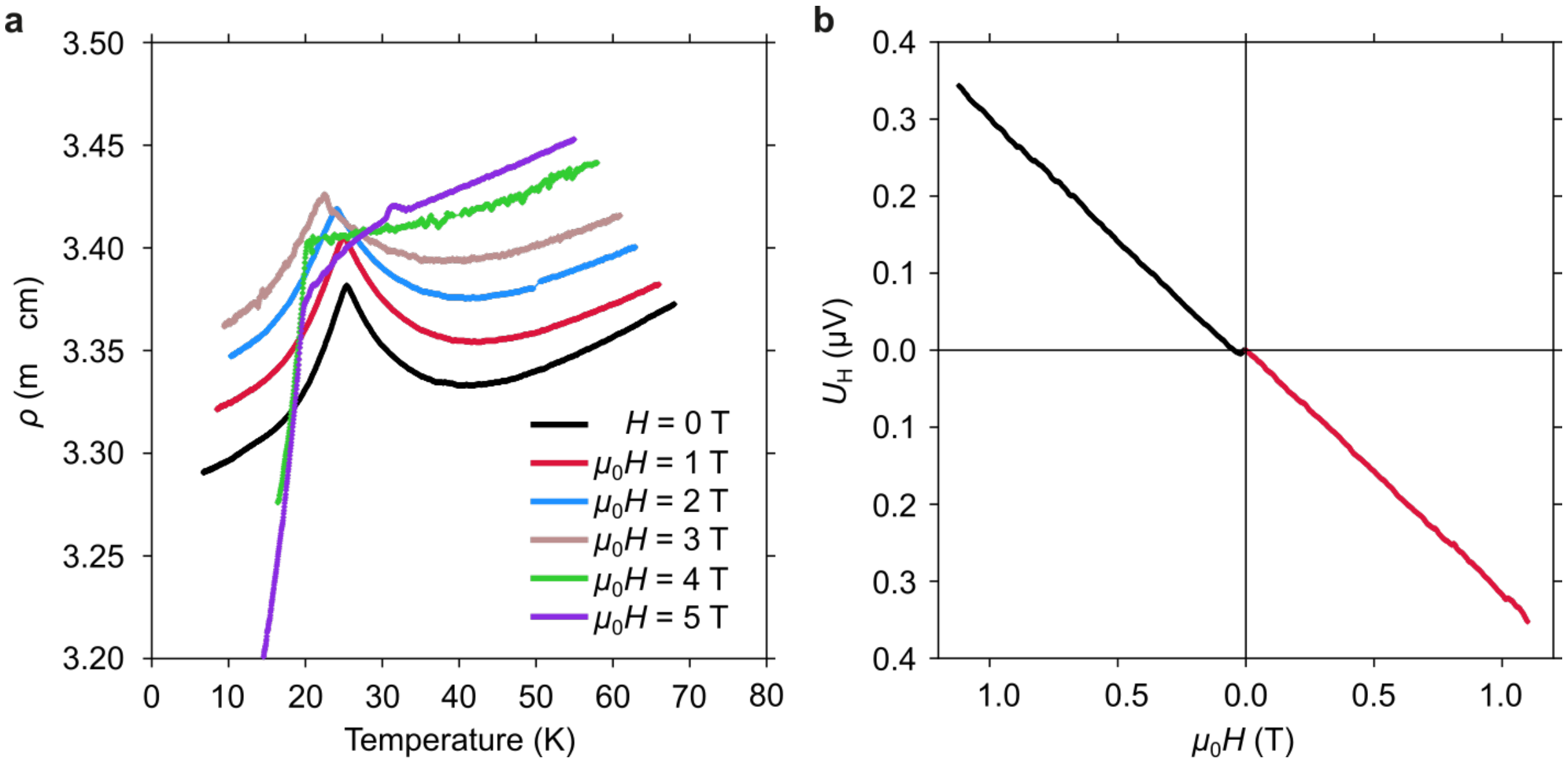

# Extended Data Fig. 4

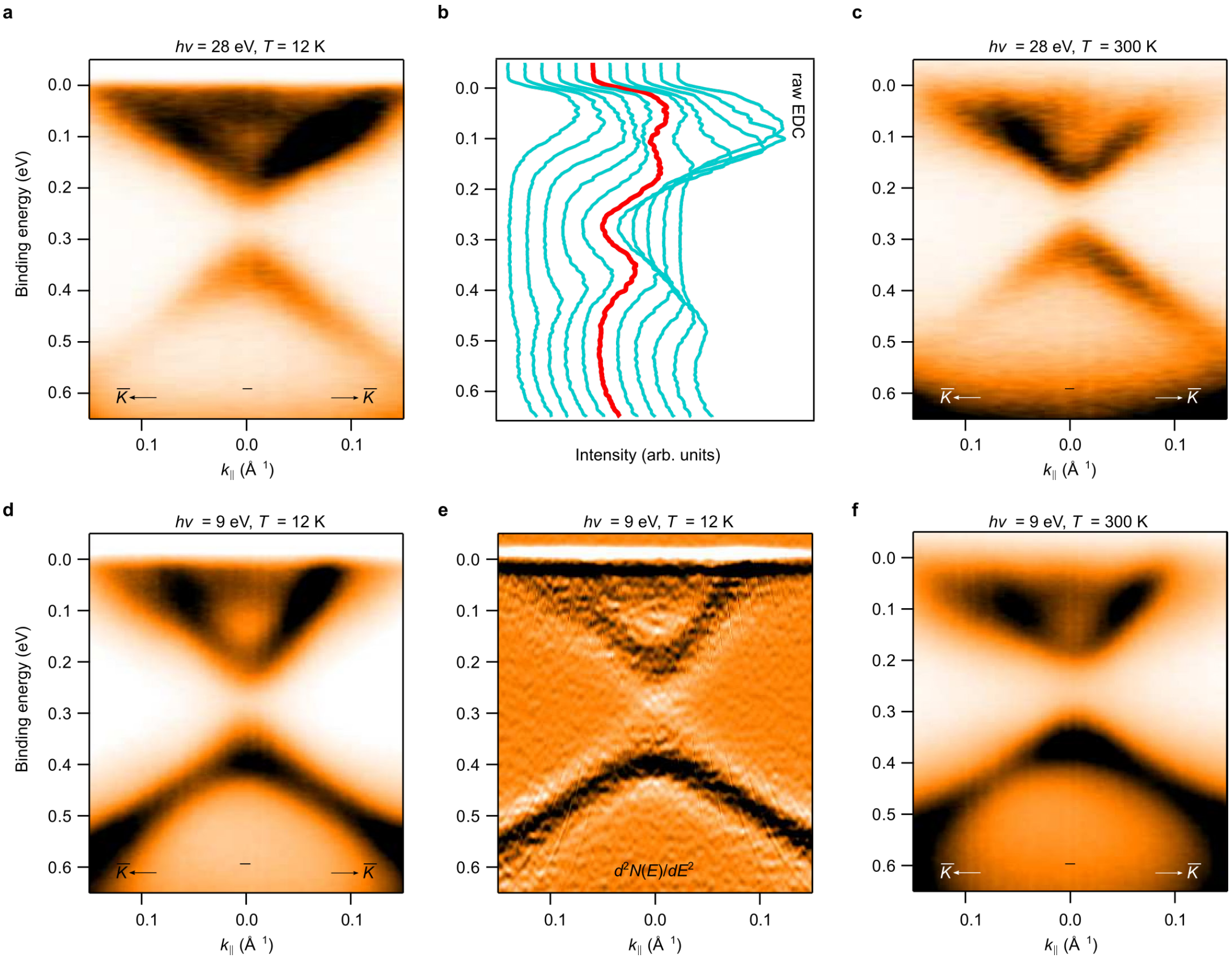

# Extended Data Fig. 5

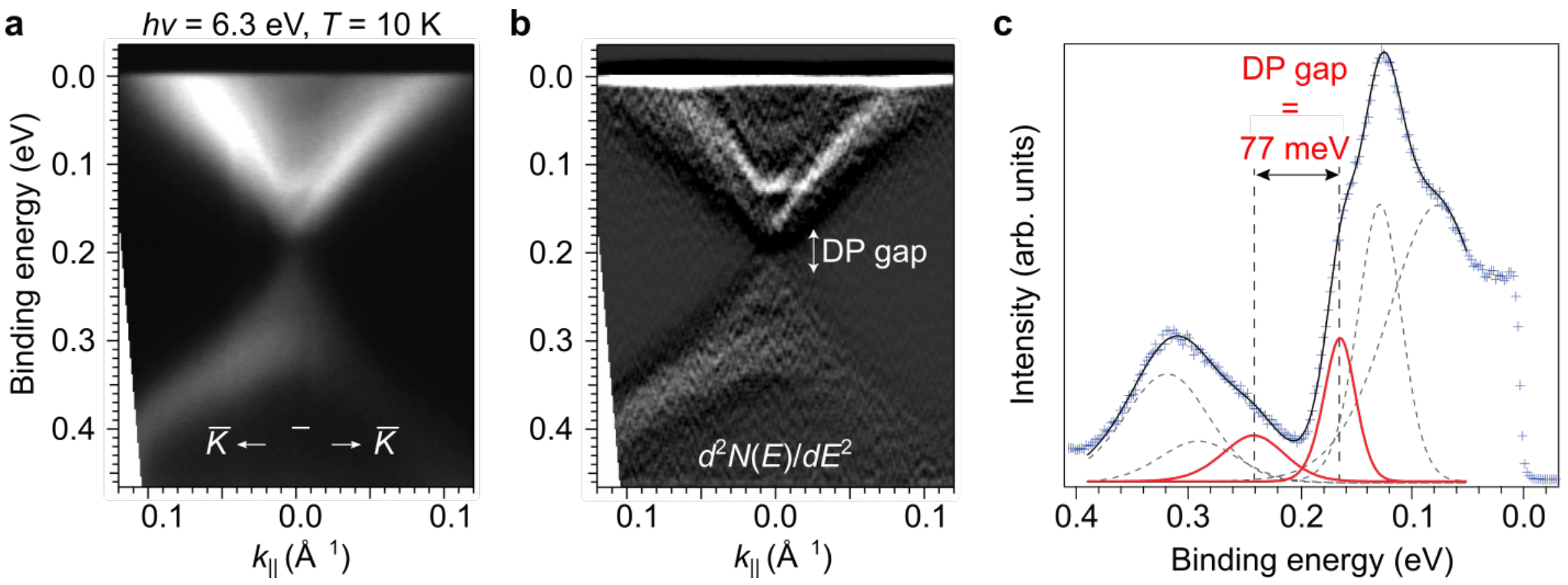

**Extended Data Fig. 6**

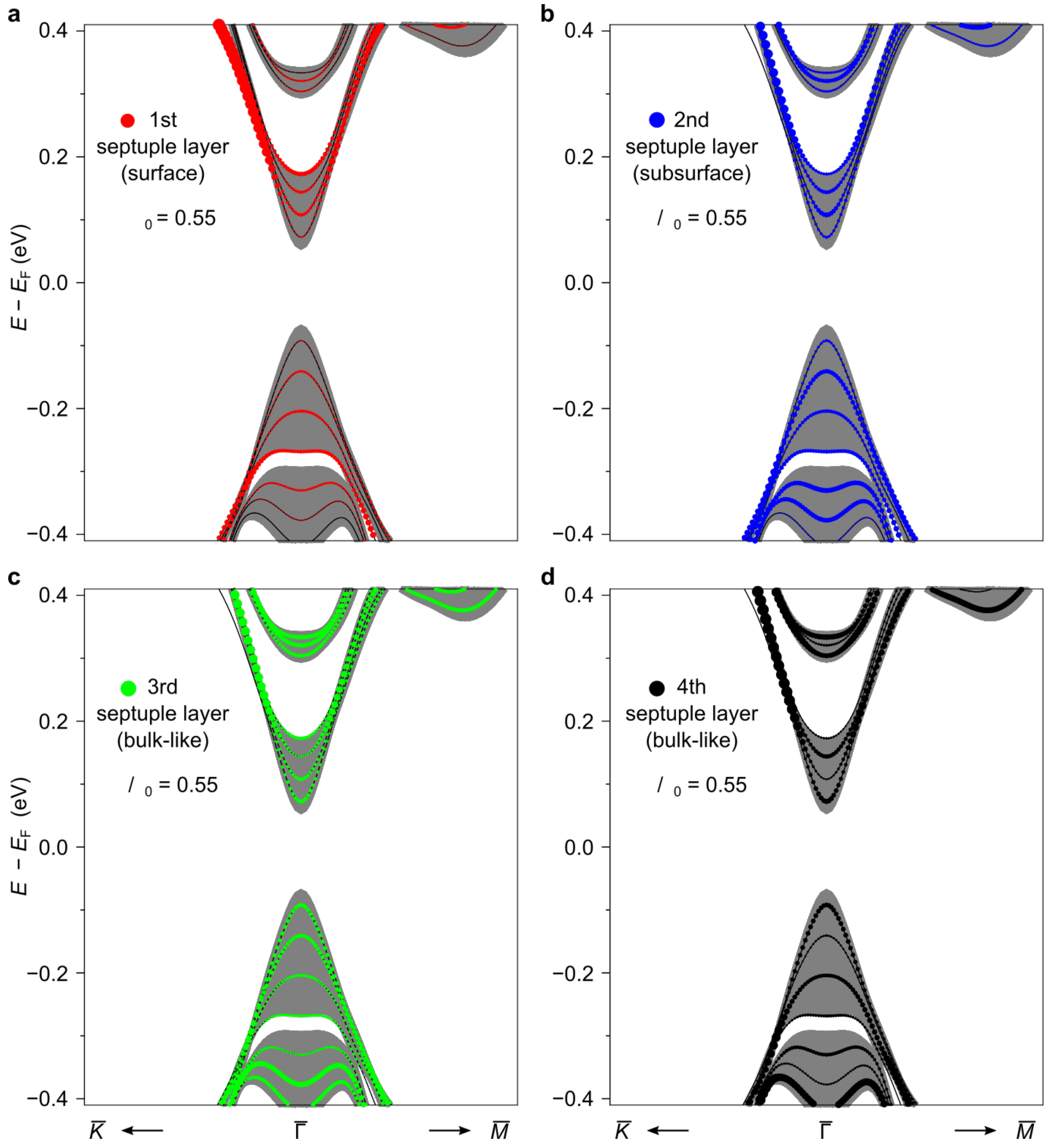

**Extended Data Fig. 7**

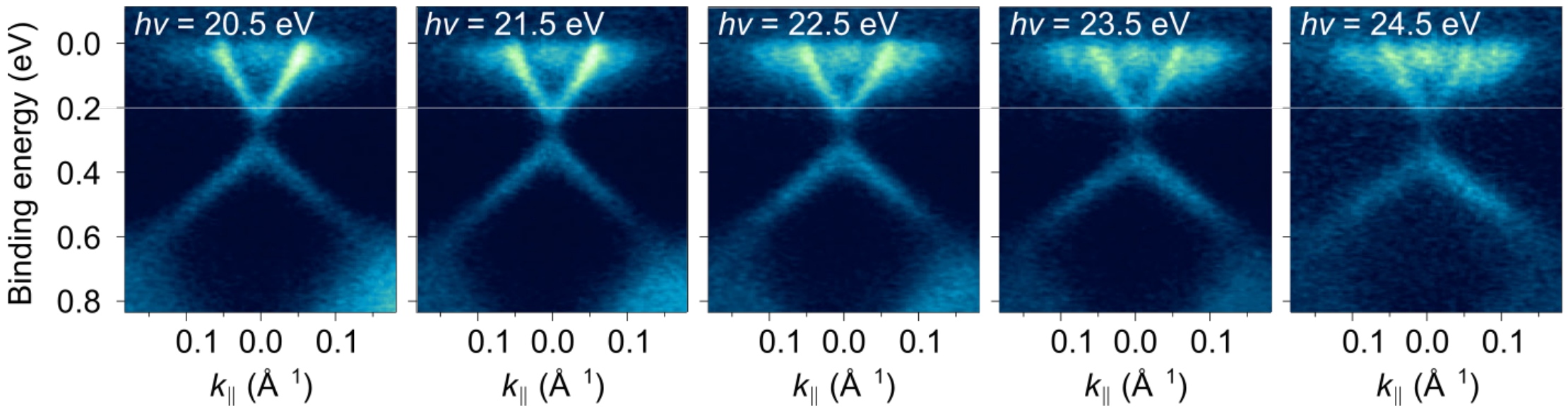

**Extended Data Fig. 8**

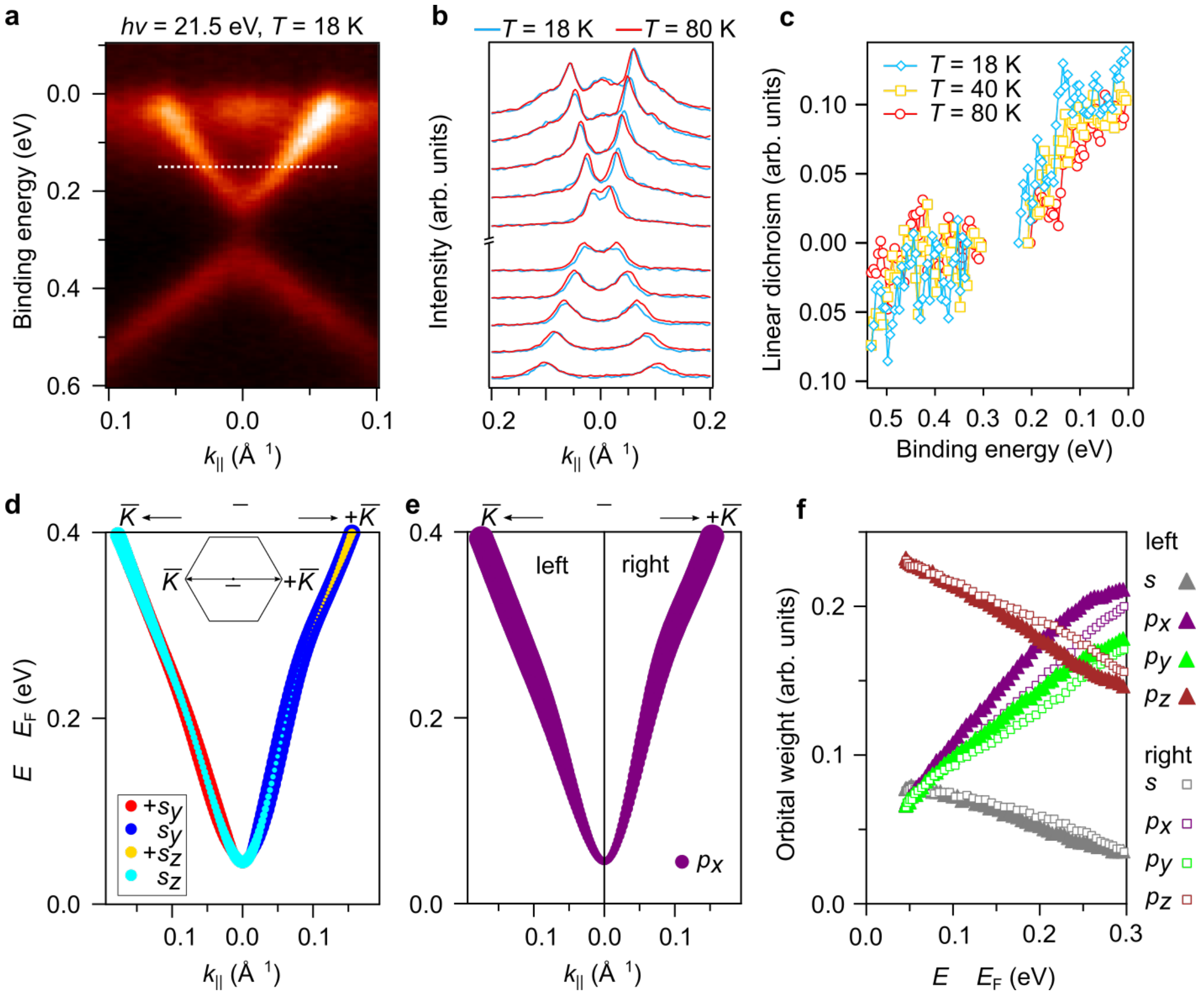

# Extended Data Fig. 9

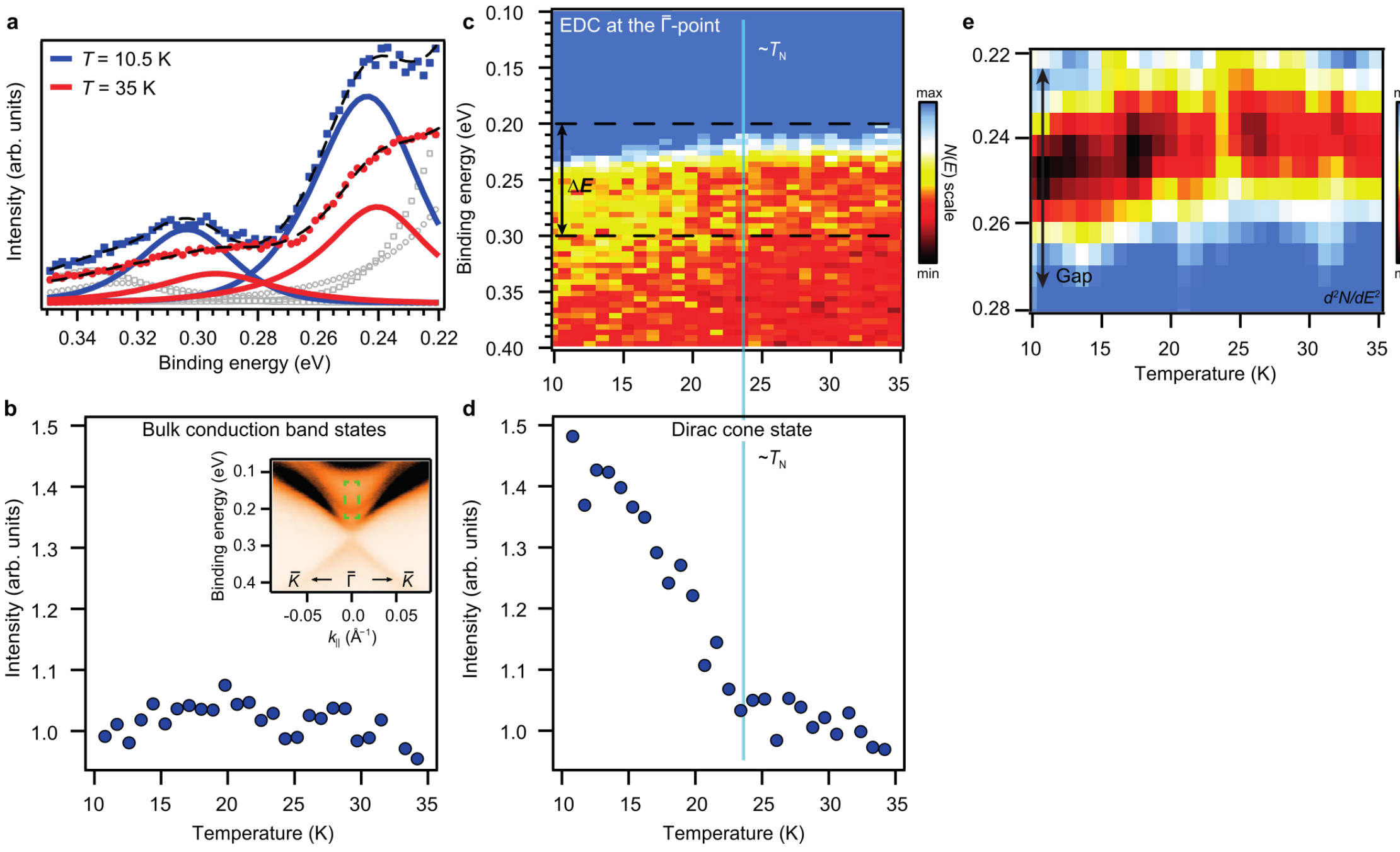

# Extended Data Fig. 10

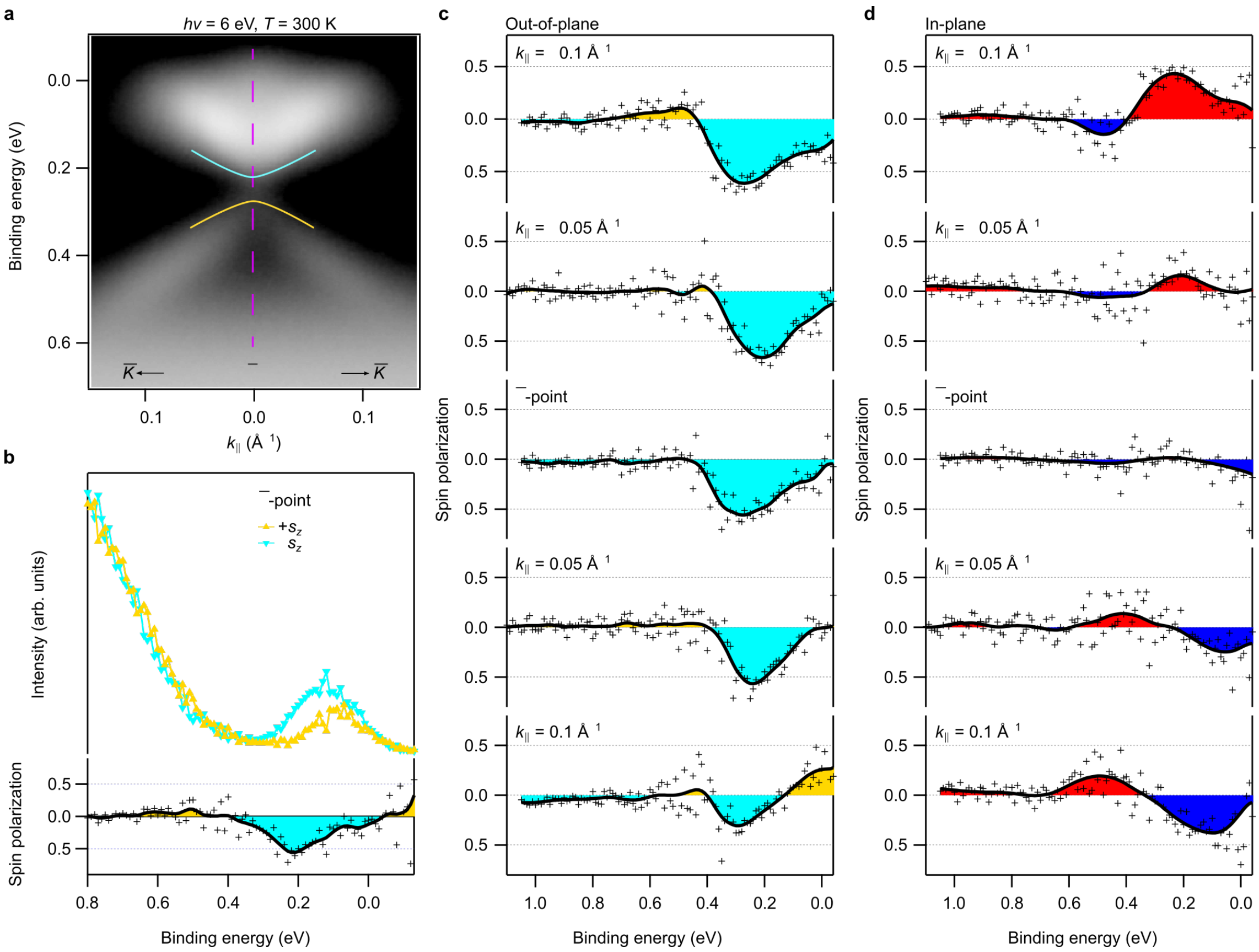